\documentclass[page-classic]{epl2} 

\usepackage{array,ragged2e,longtable}
 \usepackage{multirow}
 \usepackage{caption}
 \usepackage{subfig}
\usepackage{longtable}
\usepackage{mathtools}
\usepackage{amsmath}
\usepackage{mathtools}
\usepackage[colorlinks=true,linkcolor=blue,citecolor=blue,allcolors=blue]{hyperref}
\usepackage[top=2cm, bottom=2cm, left=2cm, right=2cm,headheight=50pt,headsep=20pt]{geometry}
\usepackage{subfig}
\usepackage{lmodern}
\usepackage{setspace}
\usepackage{url}
\usepackage[dvipsnames]{xcolor}
\usepackage{epstopdf}
\usepackage{dcolumn}
\usepackage{float} 
\usepackage{bm}
\usepackage[mathlines]{lineno}
\usepackage{flafter}
\usepackage{xcolor}
\definecolor{green78}{cmyk}{0.99,0,0.52,0}
\definecolor{dgreen}{rgb}{0.,0.6,0.}
\definecolor{turk}{RGB}{0,206,209}
\definecolor{kirmizi}{RGB}{128,0,0}
\definecolor{red1}{RGB}{220,20,60}

\DeclareGraphicsExtensions{.eps}
\usepackage{graphicx,epstopdf}
\usepackage{graphicx}

\title{The Radial Distribution Functions of Nanofluids: Molecular Dynamics Simulations}
\shorttitle{Title} 

\author{Özlem Öztürk*}

\institute{Faculty of Applied Sciences, İstanbul Aydın University, 34295 İstanbul, Türkiye.}

\abstract{Nanofluids, which are composed of insoluble, stable, and well-dispersed solid particles of nanoscale and/or subnanometer sizes suspended in a base liquid, are the next generation of liquids of today. The purpose of this paper is to investigate the one dimensional and three dimensional angle dependent radial distribution functions RDF and ARDF of polymeric nanofluids made up of nonrigid (soft) nanoparticles and a polymer melt (base fluid) using the molecular dynamics simulation approach and to search the shape stabilities by using these results. For this purpose, we use the nanoparticles of three different sizes: 28, 42, and 56 particles. We research them both within the base fluid and without this polymeric medium for instability analysis. We found that the nanoparticles with 28 atoms show the shape instability inside the base fluid when we increase the system temperature from T=1.2 to T=1.8 and hence, the structure of two concentric spherical shell of the nanoparticle breaks down and as a result the empty vacuum between these inner and the outer shells disappears. In contrast to this findings, the nanoparticles with 42 and 56 atoms show the shape stability inside the base fluid by preserving their concentric shell structures when we rise the system temperature and decrease the affinity between the nanoparticles and the base liquid medium. \\ \\
*Corresponding author: ozlemozturk@aydin.edu.tr }

\begin{document}

\maketitle

\section{Introduction}

Today's new generation liquids are 
nanofluids which compose of
insoluble, stable/well-dispersed  nanometer and/or subnanometer-sized solid
particles suspended in a base liquid \cite{Choi1995,Puliti2011}. These engineered new class nanofluids/colloidal suspensions own immense 
implementations in plenty of  scientific research fields 
ranging from medicine \cite{Dotan2016,Jones2007}, biotechnology\cite{Kavitha2018} and drug delivery\cite{Zomorodbakhsh2020,Tripathi2014} to computer hardware technologies\cite{Rafati2012} , automotive engineering\cite{Leonga2010},  nuclear engineering\cite{Hadad2010}, and microfluidics\cite{Yi2012}.

Nanofluids (nanocolloids, nanocomposites, nanosuspensions/dispersions) can be used as electro-chemical energy storage agents; in this regard for the first time reduced graphene oxide(rGO) in aqueous sulfuric acid solution was  utilized for a fast energy storage by Dubal et. al. \cite{Dubal2016} Their measurements showed that these graphene electroactive nanofluids 
exhibit high energy density and high capacitance. 

In this day and age, solar energy is an alternative renewable free energy source as compared to fossil fuels. 
Nanofluids can be employed 
as working fluids in solar collectors as an alternative of 
ordinary base liquids such as 
water and ethylene glycol (EG).
For instance, Natividade et. al. 
\cite{Natividade2019} employed 
multilayer graphene (MLG)-water nanofluid substituded for traditional liquids in the evacuated tube solar collector
equipped with parabolic concentrator. They observed the thermal efficiency enhancement by up to 76\% even at very
low MLG concentrations. 

Magnetic nanofluids \cite{Hedayatnasab2017} can be employed in hyperthermia therapies. Magnetic nanofluid hyperthermia therapy (MNFHT) used first by Gilchrist et al.  \cite{Gilchrist1957} is a therapeutic technique 
using nanofluids and acting locally for destroying cancer cells by exposing them to high enough temperatures without subjecting to the
whole human body to these temperatures. 
For example Jang et. al. \cite{Jang2017} developed Mg (magnesium) shallow doped FDA approved high biocompatible and biodegradable superparamagnetic nanoparticles $ \rm \gamma \mbox{-}   Fe_{ 2} O_{ 3}$  ( $ \rm Mg 0.13 \mbox{-}  \gamma \mbox{-}  Fe_{ 2} O_{ 3} $ nanofluids) for completely demolishing cancer cells. 

Heat exchangers are employed to transfer heat between two or more fluids at dissimilar temperatures \cite{Kakac1998}. They are used in different industries such as electric and nuclear power plants, heating, ventilation, and air-conditioning applications, 
oil refineries, and refrigeration.
Nanofluids are employed as working fluids instead of traditional liquids/suspensions in heat exchangers to hinder the erosion of the devices, to avert the blocked flow channels, to reduce the pressure drop, and to increase the heat transfer rate of the working fluid. Farajollahi et. al. \cite{Farajollahi2010} investigated the heat transfer properties of $\rm \gamma \mbox{-}Al_{2}O_{3} $/water and $ \rm TiO_{2} $/water nanofluids  in a shell and tube heat exchanger where these nanosuspensions passes through experimentally. They found that 
both liquids exhibit larger heat transfer coefficient than that of the base liquid at the same Peclet number.

Nanofluids can also be employed in refrigeration systems. Traditional refrigerants such as chlorofluorocarbons (CFCs), hydrofluorocarbons (HFCs) and  haloalkanes can give rise to ozon depletion and global warming. Thanks to these refrigerants' harmful impacts on the environment and needing much more electric power, researchers have  
developed advanced and innovative materials such as nano-refrigerants prepared via mingling the nanoparticles, conventional refrigerants and the lubricants.  Bi \textit{et. al.} \cite{Bi2008} investigated the performance of a domestic refrigerator which have $\rm TiO_{2}$ or $\rm  Al_{ 2} O_{ 3} $ nanoparticles in the working fluid experimentally. $\rm TiO_{2}$/Mineral Oil and $\rm  Al_{ 2} O_{ 3} $/Mineral Oil nanofluids were employed as a lubricant instead of Polyol-ester (POE) oil in the 1,1,1,2-tetrafluoroethane
(HFC134a) refrigerant. They showed that $\rm TiO_{2}$/Mineral Oil nanofluid causes 26.1\% less energy consumption. Babarinde \textit{et. al.}  \cite{Babarinde2020} used R600a/multi-walled carbon nanotube (MWCNT) nanoparticles-Mineral Oil   nanolubricant in place of  R134a refrigerant  in  a  domestic  refrigerator. They showed that 
this R600a/MWCNT nanofluid mixture has a better ability in removing heat and exhibits better power consumption in the refrigerator.

Especially in hot seasons, cooling systems for wind turbines are of great importance. One needs to dissipate a great deal of heat from wind tirbunes during their operation. If heat is not properly transferred from the system, this can lead to a temperature increase in the electrical and mechanical components and therefore a reduction of efficiency. And also elevated temperature can give rise to malfunctioning of the generators, and hence that results in overpriced repair costs, especially for offshore power plants \cite{DeRisi2014}. De Risi et.al. transferred the thermal load, which is mainly derived from the electric generator, by using nanofluids as a heat exchanger fluid in wind tower heat exchangers. Under steady-state condition, they observed that depending on the nanoparticle concentration and mass flow rate of nanofluids, the efficiency of the used cooling system increased by 30$\%$ because of using of nanofluids.    

Drinking water is an indispensable source of life like air for all living things on planet Earth, and therefore drinking water scarcity has always been a huge problem for humanity and continues to be so. There is a great need to find suitable ways to provide adequate drinking water for the world population. Nowadays, many countries are facing water scarcity, and people living in these countries drink and use dirty water in their daily lives. Since ancient times, people have tried to find a solution to this problem by using solar energy as a renewable energy source. For instance, it is known that in the 4th century B.C. firstly Aristotle had distilled dirty water through solar energy (solar distillation) to obtain clean drinking water \cite{Kalogirou2014}. The process of producing fresh drinking water by solar-driven desalination using solar energy is a very popular method, because  especially dry and arid regions struggling with the scarcity of fresh and clean drinking water are the areas where the sun has the most impact. Thanks to nanotechnology, an increase in the performance of solar-driven desalination systems has been observed. This is because salt water absorbs solar energy poorly, and if nanoparticles can be added onto or into the water, that causes a higher absorption of solar radiation, which increases the evaporation rate of the water \cite{Mahian2021}. Zeng et.al. \cite{Zeng2011} utilized floating $ \rm Fe_{3} \rm O_{4}/ C $ magnetic particles with 500 nm in average, which is synthesized by carbonization of poly(furfuryl alcohol) (PFA) with $ \rm Fe_{3} \rm O_{4}/ C $ nanoparticles, to increase solar energy-driven water evaporation rate. They added these particles to the surface of 3.5\% salt water 
and observed an increase at the water evaporation rate by as much as a factor of 2.3. In another study, it has been observed that the efficiency of the solar distillation system increases by 50\% when the nanofluids, which are prepared, for the first time, with sodium dodecyl sulfate (SDS) and carbon nanotubes (CNTs) are used. \cite{Gnanadason2012}. In addition to increasing the production rate of clean and salt-free water, nanoparticles also reduce carbon dioxide ($ \rm  CO_{2}$) emissions \cite{Sahota2017}.

In this article, by using the molecular dynamics simulation method, we aim to study the radial distribution functions of polymeric nanofluids consisting of nonrigid (soft) nanoparticles and a polymer melt (base fluid). As we stated above, nanofluids have diverse potential applications. They can be utilized as electrode materials, working fluids in solar collectors, wind turbines, or refrigerators, performance-enhancing agents in desalination systems, or killing agents for cancer cells. Considering all of these varied possible uses, uniform dispersion of nanoparticles within the base liquid is great importance for long-term stability of nanofluids without degredation and sedimentation. At this point, by calculating the nanoparticle-nanoparticle, the nanoparticle-monomer, and the monomer-monomer radial distribution function, we can obtain insight into the local arrangement of the nanoparticles around each other and monomers inside the polymer matrix.

In continuum-based approaches such as Finite Element Method (FEM) or Lattice Boltzmann Simulations, transport coefficients, such as dynamic (shear) viscosity coefficient, diffusion coefficient, or thermal conductivity coefficient, must be entered into the relevant equations used in these methods as external parameters, that is to say, these quantities need to be known 
in advance. In fact, these coefficients occur exactly as a direct consequence of the microscopic interactions and the quantities such as position, velocity and acceleration of a particle and they can be calculated directly by using molecular dynamics simulation method. Besides, we can calculate radial distribution functions by using molecular dynamics simulations. Actually in the macroscopic world, we obtain them from x-ray and neutron diffraction experiments, on the other hand, molecular dynamics simulations provide the positions of individual atoms over time, allowing for the straightforward computation of radial distribution functions from the resulting simulation trajectories.

\section{The Corsed-Grained Molecular Dynamics Model}

In this paper, we study the neutral (nonpolar, uncharged) bulk nanofluids composed of linear homopolymers as a base fluid and nanoparticles with different sizes. 
For this purpose, we use a generic coarse-grained particle-based  
molecular dynamics (MD) simulation 
technique with a microcanonical (NVE) (the total particle number, the system volume, and the system total internal energy are held constant) ensemble. All of the calculations in this research article is performed by using a homemade Fortran 90 Code. Such large programs contain many internal subroutines, resulting in very long and complex structures written one after another in a single file. This can make difficult to find errors in the program. As a solution, the main program is divided into many small external subprogram units. Thus, a Molecular Dynamics simulation program can be created to be easier to understand and to allow errors to be easily identified, by writing internal subprograms in separate files as external subroutines and/or functions.\\ \\
In all the equilibration and the data production runs in this study, we also utilize a double-precision (64-bit=8 byte) floating point arithmetic. The equations of motion are integrated with the velocity Verlet algorithm \cite{Verlet1967}. For the data production runs we utilize a time step $ \rm \Delta t = 0.001 \tau $ during all the simulations. The system temperature and the system pressure of the bulk nanofluid is set to $ \rm \frac{k_{B}}{\epsilon}T = 1.2$ and $ \rm P \frac{\sigma^{3}}{ \epsilon} = 0.0 $, respectively in all the simulations as well. We simulate the model systems that
are made up of nanoparticles composed of 28, 42, and 56 atoms inside the surrounding polymer melt of 5600, 8400, and 11200 monomers, respectively, via MD simulations. The number densities corresponding to the zero-pressure of three different systems with three different nanoparticles of 28, 42, and 56 atoms are set to $\rm \rho \sigma^{3} =0.815, \; 0.820, \;0.825 $, respectively. Interaction potential energy between any two of the atoms made up of 
our system is central, spherically symmetric (independent of polar and azimuthal angle, dependent only on the distance between two atoms), and pairwise additive. This last phrase means that the interaction between any two particles is \textit{unaffected} by all the other particles. A 12-6 Lennard-Jones (LJ) interaction potential energy function between a pair of uncharged atoms and/or molecules is given by
\begin{equation}
\begin{gathered}
\rm {V}_{\rm LJ}= \rm 
4 \epsilon \left [  \left ( \frac{ \sigma }{r} \right )^{12}
-  \left ( \frac{ \sigma }{r} \right )^6  \right ]
\end{gathered} 
\label{eq:LJ}
\end{equation}
where r is the central distance 
between atom pairs. $ \sigma $ is the distance 
at which the potential energy is zero, and also gives the characteristic/approximate particle diameter. 
$ \epsilon $ is the depth of the potential well and also controls the strength of the interaction. The 12-6 Lennard-Jones potential consists of two terms. The positive (repulsive) short-range term 
is responsible for excluded volume effects of the atoms: it describes interactions resulting from Pauli's exclusion principle. Thanks to this term, a strong repulsive force occurs between atom and/or molecule pairs when they get too close to each other. And the negative (attractive) long-range term represents the van der Waals interactions and it allows particles at long distances to attract each other, allowing, for example, a system in the gas phase to condense into a liquid phase. A 13-7 Lennard-Jones force, $ \vec {\rm F}_{\rm LJ}=- \vec{\nabla}\rm V_{LJ} $ , is given by as follows:
\begin{equation}
\begin{gathered}
\vec {\rm F}_{\rm LJ}= \rm 
\frac {24 \epsilon}{\sigma} \left [ 2 \left( \frac{ \sigma }{r} \right )^{13}
-  \left ( \frac{ \sigma }{r} \right )^7  \right ] \hat r
\end{gathered} 
\label{eq:LJforce}
\end{equation}
where $\rm \hat r$ is the unit vector.

\begin{figure}[H]
	\centering
	\includegraphics[height=6cm]{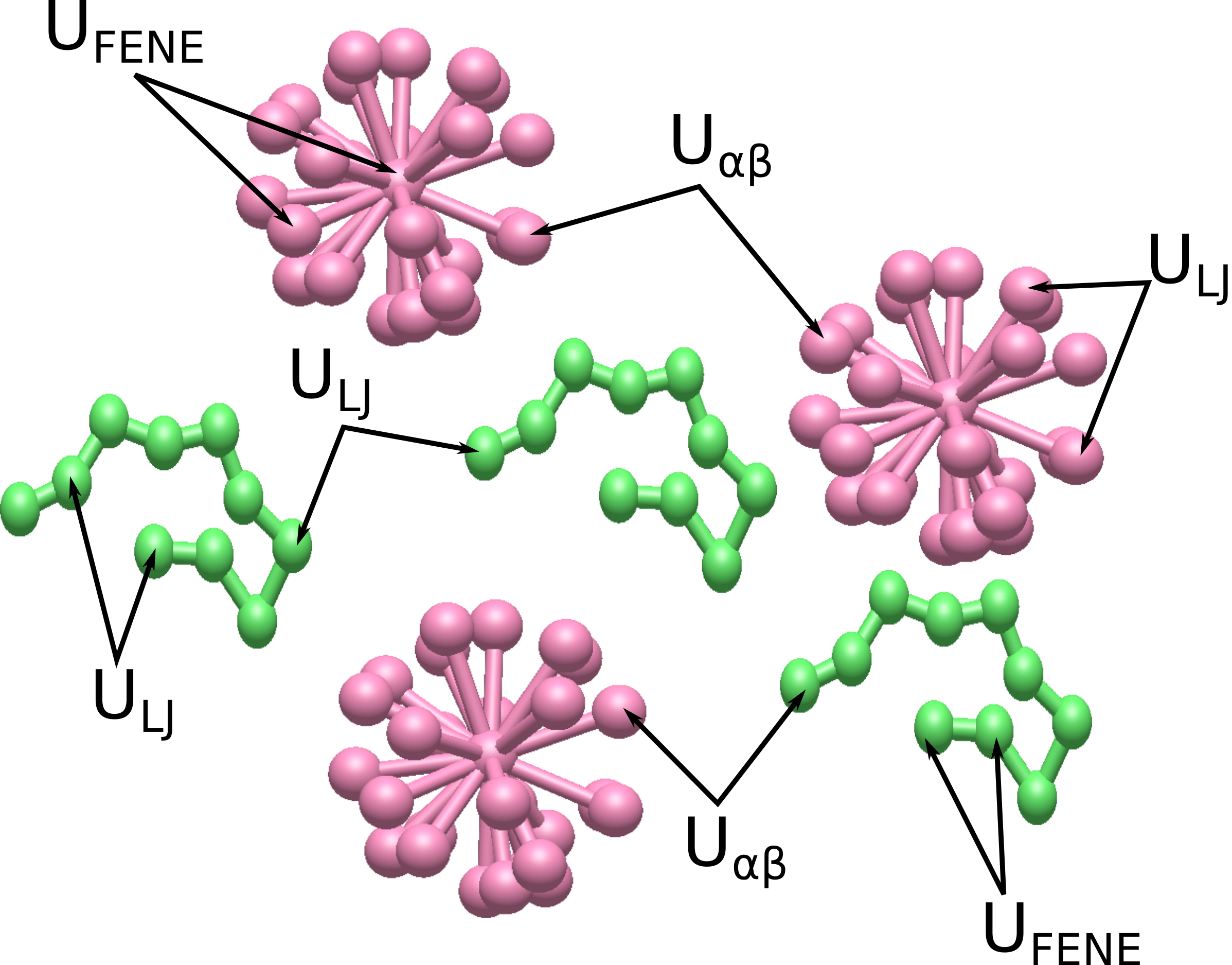}
	\caption{An illustration of the interactions inside the bulk nanofluid systems of 28, 42, and 56 particles. In the figure, three nanoparticles (each one of 28 atoms) are shown in pink colour. And also three polymer chains (each one of 10 monomers) are seen in green colour. As you can see, there are seven different interaction types based on where they occur. There are four intramolecular interactions and three intermolecular interactions. In this figure, $\rm U_{FENE}$, $\rm U_{LJ}$, and $\rm U_{\alpha \beta}$ represents FENE potential energy function, 12-6 LJ potential energy, and \textit{modified}-12-6 LJ potential energy respectively.}
	\label{fgr:interactions}
\end{figure}

There are seven different interactions in the bulk nanofluid (See Figure \ref{fgr:interactions}): The adjacent monomer pairs in a same polymer chain interact via \textit{finitely extensible nonlinear elastic} (FENE) potential energy \cite{Grest1986,Kremer1990}. The monomer pairs in the same polymer interacts via 12-6 LJ potential energy function, and the two monomers from two different polymer chains also interacts via 12-6 LJ potential as well. Between the central atom  and one of the other particles in a nanoparticle interact via FENE potential energy. The atom pairs in a same nanoparticle interacts via 12-6 LJ potential energy function, and the two particles from two different nanoparticles also interacts via \textit{modified}-12-6 LJ potential. Finally, a nanoparticle atom and a monomer interacts via \textit{modified}-12-6 LJ potential as well (See the Eqs\eqref{eq:LJ}-\eqref{eq:Ffene} and Eqs\eqref{eq:UmCnpa}-\eqref{eq:ModfULJ}).
\subsection{Monomer-Monomer Interaction} 
The base fluid used in this study consists of neutral, solvent-free macromolecules, and also it is the liquid composed of \textit{linear} polymer chains. Each polymer chain consists of $\rm 10$ homogeneous monomers in total. Maximum $\rm 11200 $ monomers has been used to form liquid polymer matrix. In this research, the two-body, bonded (intramolecular) interactions between any two monomers inside a polymer chain or nonbonded (intermolecular) interactions between the two monomers in any two distinct polymer chains are modelled by a truncated and shifted (i.e., continuous) 12-6 Lennard-Jones potential energy function (see Figure \ref{fgr:interactions} and Eqs.\ref{eq:LJ}-\ref{eq:LJforce}) is
as the following:
\begin{equation}
\rm {U}_{\rm LJ} (\rm r)= \rm \left \lbrace
\begin{array}{cc} \rm {V}_{\rm LJ} (\rm r) - \rm {V}_{\rm LJ} (\rm r_{\rm c})
& \rm \text{for} \;  r < r_{c}    \\
0 \rm  & \rm \text{for} \; r  \ge  r_{c} 
\end{array}
\right.
\label{eq:ULJa}
\end{equation}
where 
\begin{equation}
  \rm {V}_{\rm LJ}(\rm r_{\rm c})=-\frac{127 \epsilon}{4096}.
   \label{eq:ULJb}
\end{equation}  
The cut-off distance is 
$ \rm r_c=2 \times 2^{1/6}\sigma $. This distance 
causes to happen liquid-vapor 
phase separation and droplet formation for appropriate 
thermodynamics conditions below the theta temperature, $ \rm  \Theta =3.3 \epsilon /k_{B} $ and allows us to model poor solvent conditions
\cite{MacDowell2000, Müller2001,Pastorino2007,Servantie2008a, Servantie2008b, Pastorino2015}. r is the central distance 
between particle pairs. $ \sigma $ is the distance 
at which the potential is $\rm {U}_{\rm LJ} (\rm \sigma)=- \rm {V}_{\rm LJ}(\rm r_{\rm c})\approx -0.03 \epsilon  $ ($\rm {V}_{\rm LJ}(\rm \sigma)=0  $), and also gives the characteristic/approximate particle diameter. 
$- \epsilon $ and $- 0.97 \epsilon $ is the depth (i.e., the potential energy reaches its minimum value) of the potential well of $\rm {V}_{\rm LJ}(\rm r_{\rm min})$ and $\rm {U}_{\rm LJ}(\rm r_{\rm min})$ with $\rm r_{\rm min}=2^{1/6} \sigma $ respectively, and they also control the attractive strength of the interaction.  The Lennard-Jones parameters $ \sigma $ and $\epsilon $ defines units of length and 
energy, respectively in this study. All the monomers have identical mass, m, and it defines unit of mass. In addition,  $ \rm \tau=\sqrt{m\sigma^{2}/ \epsilon} $ defines unit of time in all the simulations. Because all the equations used in the simulations are nondimensionalized, $\rm  \sigma $,  $\rm \epsilon$, $\rm m$, and $\rm \tau$ are set to be equal to one for all the MD simulations in this study.\\ 
  The bonded (intramolecular) interactions between the adjacent monomer pair of an uncharged polymer chain (See Figure \ref{fgr:interactions}) are modelled by the FENE
 potentail energy function \cite{Grest1986,Kremer1990}, 
 \begin{equation}
 \rm {U}_{\rm FENE}=\rm \left\lbrace
 \begin{array}{cc} \rm
 -  \frac{1}{2} k {R_{0}}^{2}  \ln  \left[ 1  -  \left  
 (  \frac{r}{R_{0} } \right )^2  \right ]  & \rm  \text{for} \; r< R_{0}   \\ 
 \infty  & \rm \text{for} \;r \ge R_{0} 
 \end{array}
 \right.
 \label{eq:fene}
 \end{equation}
 where the maximum covalent bond length between the adjacent monomer pair is $ \rm R_{0}=1.5\sigma $, and 
 the spring constant is $\rm k=30 \epsilon /\sigma^{2}$.
 $ \rm r $ is the distance between the centers of the particles forming the consecutive monomer pair. This FENE potential keeps the bonds between any adjacent two monomers on the same linear polymer chain of 10 monomers together. Kremer and Grest \cite{Grest1986,Kremer1990} stated in their study that these numerical values of $ \rm R_{0}$ and $ \rm k $ prevent a single polymer chain to cut itself and/or the other polymer chain.\\
 FENE force is given by
  \begin{equation}
 \vec {\rm F}_{\rm FENE}=
   \rm  \frac{kr}{ \left  
 	(  \frac{r}{R_{0} } \right )^2-1} \hat r 
   \label{eq:Ffene}
 \end{equation} 
   In Eq.\ref{eq:Ffene},  FENE force, $ \vec {\rm F}_{\rm FENE}=- \vec {\nabla} \rm {U}_{\rm FENE}   $, is an attractive force for $ \rm (0,R _{0})$ interval, and $ \rm r \rightarrow R _{0}^{-}$, $ \vec {\rm F}_{\rm FENE}  \rm  \rightarrow -\infty $ and $ \rm r \rightarrow R _{0}^{+}$, $ \vec {\rm F}_{\rm FENE} \rm  \rightarrow +\infty $. The FENE force is given in the Figure \ref{fgr:FENEforce}  below:
   \begin{figure}[H]
  	\centering
  	\includegraphics[height=7cm]{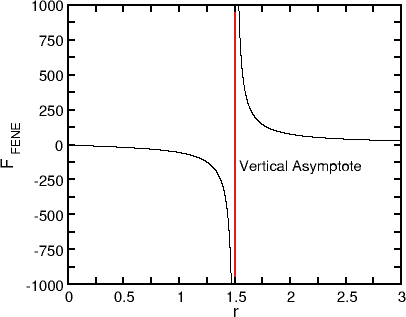}
  	\caption{For the given values of the parameters $ \rm k $ and $ \rm R_{0} $, the variation of the component of  $\vec{\rm F}_{\rm FENE} $ function in the direction of unit vector $ \rm \hat r $ with the distance r. Red vertical line represents the vertical asymptote at the point r=1.5.}
  	\label{fgr:FENEforce}
  \end{figure}
     When considering the limit situation, $  \lim_{\rm r \rightarrow \rm R _{0}^{+}} \vec {\rm F}_{\rm FENE}=\rm  +\infty $, the distance, r, in Eq.\ref{eq:Ffene}, can be greater than $\rm  R _{0}$ during the equilibration run, hence $\vec{\rm F}_{\rm FENE} $ can be repulsive, therefore, just in case, we have to check whether or not it is greater than $\rm  R _{0}$ throughout the molecular dynamics simulations lest the bonds between any consecutive two monomers on a same polymer chain do not break. An appropriate neighbour list for FENE interaction also must be constructed before starting the simulations depending on whether we use Third Law of Newton or not as well.
     \subsection{Nanoparticle-Nanoparticle Interaction} 
\begin{figure}[H]
 	\centering
 	\includegraphics[height=7cm]{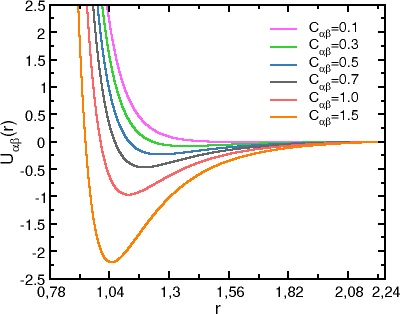}
 	\caption{For six different $\rm  C_{\alpha \beta} $ coefficients, \textit{modified}, truncated and shifted Lennard-Jones (12,6) potential energy, $\rm U_{\alpha \beta}(r) $, as a function of the central distance r between two particles is shown. For $\rm  C_{\alpha \beta}=0.1 $, the minimum of the $\rm U_{\alpha \beta}(r) $ is approximately equal to $ -0.0071 $ at $\rm r=1.6$.}	\label{fgr:UmCnp}
 \end{figure}

An atom-pair from the two different nanoparticles interacts with a \textit{modified}, truncated, and shifted 12-6 Lennard-Jones potential energy function \cite{Barrat1999} (See Figure \ref{fgr:interactions}):
 \begin{equation}
 	\rm {U}_{\rm \alpha \beta} (\rm r)= \rm \left \lbrace
 	\begin{array}{cc} \rm {U}_{\rm \alpha \beta} (\rm r) - \rm {U}_{\rm \alpha \beta} (\rm r_{\rm c})
 		& \rm \text{for} \;  r < r_{c}    \\
 		0 \rm  & \rm \text{for} \; r  \ge  r_{c} 
 	\end{array}
 	\right.
 	\label{eq:UmCnpa}
 \end{equation}
 \begin{equation}
 	\rm U_{\rm \alpha \beta} (\rm r)= \rm 
 	4 \epsilon  \left [  \left ( \frac{ \sigma}{r} \right )^{12} 
 	- C_{\rm \alpha \beta}  \left ( \frac{ \sigma}{r} \right )^6  \right ]
 	\label{eq:ModfULJ}
 \end{equation}
 where $  \alpha , \beta $ = m, n denotes the interaction which is for a monomer (m) or a nanoparticle atom (n). The dimensionless coefficient (positive real number) $ \rm C_{\alpha \beta}$ \cite{Barrat1999}  rules the attractive part of the potential $\rm U_{\rm \alpha \beta} (\rm r) $. The larger the $ \rm C_{nn}$ is, the stronger the affinity between the nanoparticles is. In order to be able to have a well dispersed nanofluid, the aggregation of  nanoparticles has to be averted. That corresponds to the lack of strong affinity between the nanoparticles (the dispersed phase) and the polymer melt (i.e., the liquid phase/the dispersion medium in which they are dispersed) and accordingly to the more lyophilic nanoparticles. We discovered that for nanoparticle-nanoparticle interactions, $ \rm C_{nn}=0.1 $ (see Figure \ref{fgr:UmCnp}) guarantees a well dispersed nanofluid (nonflocculating nanoparticles) and we fixed it to this value throughout the simulations. 

  \subsection{Construction of a Nanoparticle} 
\begin{figure}[H]
  	\centering
 	\includegraphics[height=3cm]{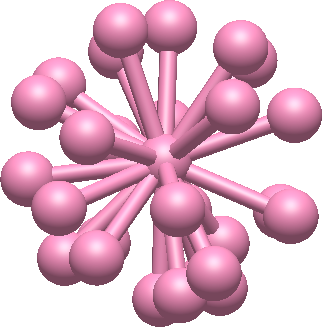}
 	\caption{The visualization of a spherical nanoparticle with 28 atoms created by using the Algorithm in Table \ref{tab:nanoparticle}. The 27 bonds emerging from the central atom are seen. For visualization, we use Avogadro 2 \cite{Hanwell2012}. }
 	\label{fgr:nanoparticle28}
 	\end{figure}

In our spherical nanoparticle model systems with two homocentric shells, we analyze nanoparticles of three different sizes with  28, 42, and 56 atoms in total. Each nanoparticle consists of two concentric spherical shells and a central atom, which is surrounded by the other atoms in the same nanoparticle, at the center of these two concentric shells of the inner radius $ \rm R_{in}$ and the outer radius $ \rm R_{out}$. Each nanoparticle with two homocentric shells are made up of identical atoms, and each of these atoms has the same mass as the polymer atoms. 
  These spherical nanoparticles with different sizes are created by writing a homemade Fortran 90 code in this study. In this process, first of all, we use spherical coordinates $ (\rm r, \theta,\phi) $ to construct a nanoparticle in a spherical geometry and then convert these coordinates to cartesian coordinates to be able to obtain the initial configurations of the nanoparticles in a cartesian coordinate system, as we do all of the simulations in this coordinate system throughout this article. The first indexed atom is placed at the center (origin). We fix the radial distance from the central atom, $\rm r=\rm r_{0} $,  to $ \rm 1.50, \; 1.60, \; 1.69$ for the nanoparticles of $\rm 28, \; 42, \; and\; 56\; $ atoms, respectively, and $\rm d_{0}=0.55 $. In here, $ \rm d_{0}$ represents the minimum distance between any two atoms inside a nanoparticle. The distance between any two particles is set to be greater than or equal  to 0.55 during the execution. Adjusting the minimum distances in this way makes the nanoparticle equilibration stage relatively easy. And then, after placing the first indexed atom at (0,0,0) point, the other atoms are stowed randomly on a spherical surface by changing only $ \rm \theta $ and $\rm \phi $ coordinates during the execution. In order to be able to provide randomness, we use \textbf{ran2}\cite{Press2000}, which is a uniform random number generator. These uniformly distributed random numbers are between 0.0 and 1.0 excluding endpoints, and their mean and the standard deviation are equal to  $0.5$ and $\rm 1/\sqrt{12} $  respectively. Assume that we choose a random real number, p, from an interval $\rm (a,b)$. In order to be able to convert the interval $\rm (a,b)$ to the interval $\rm (c,d)$, we can use the transformation as the following,
  \begin{equation}
  	\rm p \longmapsto p 
  	\left( \frac{d-c}{b-a}  \right) +
  	\left( \frac{bc-da}{b-a}  \right)
  	 \label{eq:convert1}
  \end{equation}
  Since random number generator, \textbf{ran2}\cite{Press2000}
  generates a real number in the interval $\rm (0,1) $, we have to use the transformation $\rm  ( for \; a=0, \; b=1) $,
  \begin{equation}
  	\rm p \longmapsto p 
  	(d-c) +c  	 
  	\label{eq:convert2}
  \end{equation}
  and for the polar angle $\rm \theta $ and the azimuthal angle $\rm \phi $, we have to take the intervals $\rm (0,\pi) $ and $\rm (0, 2\pi) $, and use the corresponding transformations,
  \begin{equation}
  	\rm p \longmapsto p \times
  	\pi
  \end{equation}
  \begin{equation}
  	\rm p \longmapsto p \times
  	2\pi \\
  \end{equation}
  respectively.
   The algorithm to create a nanoparticle is stated in Table~\ref{tab:nanoparticle}:
  \begin{table}[H]
  	\centering\hrulefill
  	\caption{An algorithm to construct a spherical nanomolecule of different sizes.}
  	\hrulefill
  	\begin{enumerate}
  		\item Assign the value of the effective radial distance, $ \rm r_{0} $,  and the effective minimum central distance between the particles, $ \rm d_{0} $,
  		and the first indexed-atom's 3D coordinates, \:
  		\item Looping over all atoms in a nanoparticle \:
  		\item Call \textbf{ran2} for polar angle, \:
  		\item Call \textbf{ran2} for azimuthal angle, \:
  		\item Convert \textbf{ran2}'s uniform random number interval from $ (0,1) $ to $\rm (0,\pi) $ and $\rm (0,2\pi) $ for polar and azimuthal angles, respectively,\:
  		\item Transform spherical coordinates to cartesian coordinates:\\
  		$ \rm x=rsin\theta cos \phi$ \\ 
  		$ \rm y=rsin\theta sin \phi$ \\
  		$ \rm z=rcos \phi$ \\
  		\item if the distance between any two atoms is less than $\rm d_{0}$, go to 3, otherwise go to 8, 
  		\item Go to 2,
  	\end{enumerate}
  	\hrulefill
  	\label{tab:nanoparticle}
  \end{table}
 In addition, the visualization of a spherical nanoparticle with 28 atoms just after created by using the Algorithm in Table \ref{tab:nanoparticle} is shown in Figure \ref{fgr:nanoparticle28}.  
 	\subsection{Equilibration of a Nanoparticle} 
\begin{figure}[H]
    	\centering
    	\subfloat[][\centering 28 particles] {{\includegraphics[width=2.1cm]{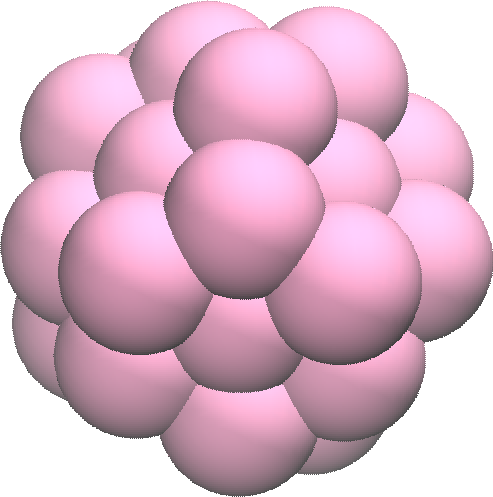} }} \hspace{1cm}
    	\subfloat[][\centering 42 particles] {{\includegraphics[width=2.6cm]{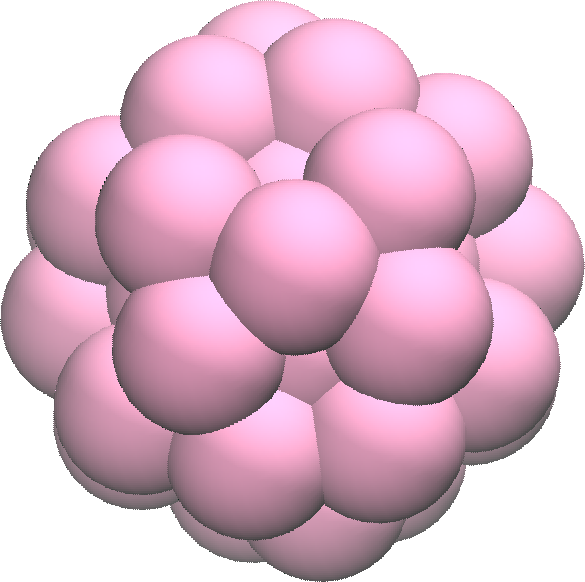} }}  \hspace{1cm}
    	\subfloat[][\centering 56 particles] {{\includegraphics[width=2.8cm]{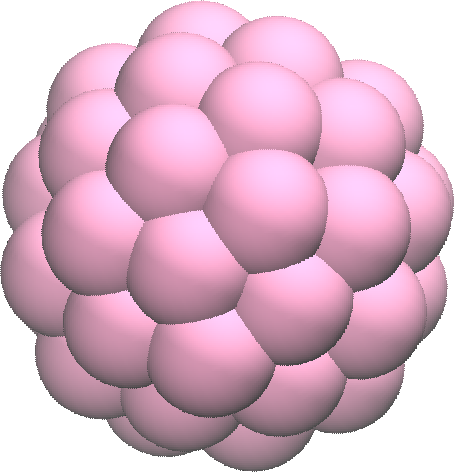} }}%
    	\caption{Side by side visualization of  spherical nanoparticles of 28 (a), 42 (b), 56 (c) atoms just after the production run of $ \rm 10000 \tau $ simulation time is shown in this figure. For visualization purporse, we use VMD \cite{Humphrey1996}. }%
    	\label{fig:28-42-56pars}%
    \end{figure}

    We equilibrate the nanoparticle systems with three different sizes in the NVE ensemble before inserting the nanoparticles into the polymeric dispersion medium. At thermodynamic equilibrium, the velocity distribution of the particles has to obey the Maxwellian distribution. Hence, by using a uniform random number generator, such as \textbf{ran1} \cite{Press2000}, and the function converting the uniform distribution to a Gaussian distribution,\textbf{ gasdev}\cite{Press2000}, we give the Gaussian distribution for the initial velocities of the nanoparticles for the desired temperature. 
    We place the center of mass of the each nanoparticle at the origin of the simulation box. At the beginning of the each simulation, we subtract the center of mass velocity from the initial velocities for once, therefore we obtain stationary nanoparticles throughout the equilibration runs. In this spherical nanoparticle model systems with three different sizes, any two atoms in a nanoparticle interact with 12-6 Lennard-Jones potential energy (See Eqs.\eqref{eq:LJ}-\eqref{eq:ULJb} and Figure \ref{fgr:interactions}). We utilize the same cut-off distance used for the polymer matrix in the potential energy function. Besides the Lennard-Jones type of interactions between two particles inside a nanoparticle as stated above, the central atom in a spherical nanoparticle of 28, 42, and 56 particles interacts the other 27, 41, and 55 atoms with FENE potential energy for $\rm  R_{0}=1.52,\; 1.62,\; and \;1.71$ values, respectively, and also the same spring constant $\rm  k=30 \epsilon/\sigma^{2} $ value for the polymer matrix (See Figure \ref{fgr:interactions} and Eqs.\ref{eq:fene}-\ref{eq:Ffene}) as well. This type of FENE interaction and these atom numbers (28, 42, and , 56) causes approximately spherical nanoparticles of two concentric shells in the absence of a bulk polymer liquid. To a FENE neighbour list construction, for instance, for N nanoparticles with 28 atoms, there are $\rm N\times28 $ particles in total, therefore we consider that there are 27 neighbours of each central atom and we constract a FENE neighbour list for MD simulation according to this fact: The $\rm 1^{st} $ indexed central atom's neighbours in the $\rm 1^{st} $ nanoparticle are 2, 3, 4, ... ,27, 28, and the $\rm 29^{th} $ indexed central atom's neighbours in the $\rm 2^{nd} $ nanoparticle are 30, 31, 32, ... , 55, 56, and so on. By using a FENE neighbour list, we can calculate FENE force, $ \rm \overrightarrow{\rm F}_{FENE}= -\overrightarrow{\nabla} U_{FENE}$, and $ \rm U_{FENE}$ potential energy for each nanoparticle as we mentioned just above. We use periodic boundary conditions in x, y, and z directions. One of the side lengths of the cubic simulation boxes is equal to $ \rm  65.42 \sigma, \; 74.89 \sigma, \; and \;  82.43 \sigma$ for nanoparticles with 28, 42, and 56 atoms,  respectively, and each of these numbers corresponds to number density, $ \rm \rho_{par} \sigma^{3} =0.0001 $. In order to bring the initial configurations of the nanoparticles, which we  calculate in the previous section \textit{"Construction of a Nanoparticle"}, into the equilibrium in the NVE ensemble, initially we utilize  the time step of $ \rm \Delta t=0.0001 \tau $ because of the high initial potential energy of the nanoparticles. By using this time step, we perform the velocity rescaling at every 10 simulation time step for 10 million simulation time steps in total. After this stage, by starting the simulation with the velocities at which the previous simulation left off, and  by taking $ \rm \Delta t=0.001 \tau $, again we perform the velocity rescaling at every 10 simulation time step for 10 million simulation time steps in total to reach the desired system temperature $ \rm (k_{B} / \epsilon) T  =1.2 $. After this phase, a production run can be performed for data collection. For data collection stage, we always use the time step $\rm \Delta t=0.001 \tau $. After collecting data for 10 million simulation time steps in total, the equilibrated nanoparticles of 28, 42, and 56 particles that we use for all MD simulations are shown in Figure \ref{fig:28-42-56pars}.

\begin{figure}[H]
  	\centering
  	\includegraphics[height=4.2cm]{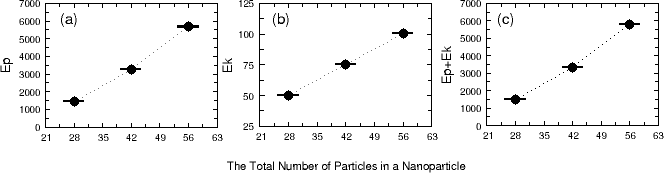}
  	\caption{The Figures show the graph of Ep (a), Ek (b), and Ep+Ek (c) versus total number of atoms in a nanoparticle. Dotted lines are used to guide to the eye.}
  	\label{fgr:nanoparenergy}
  \end{figure}

\begin{figure}[H]
 	\centering
 	\includegraphics [width=13.cm]{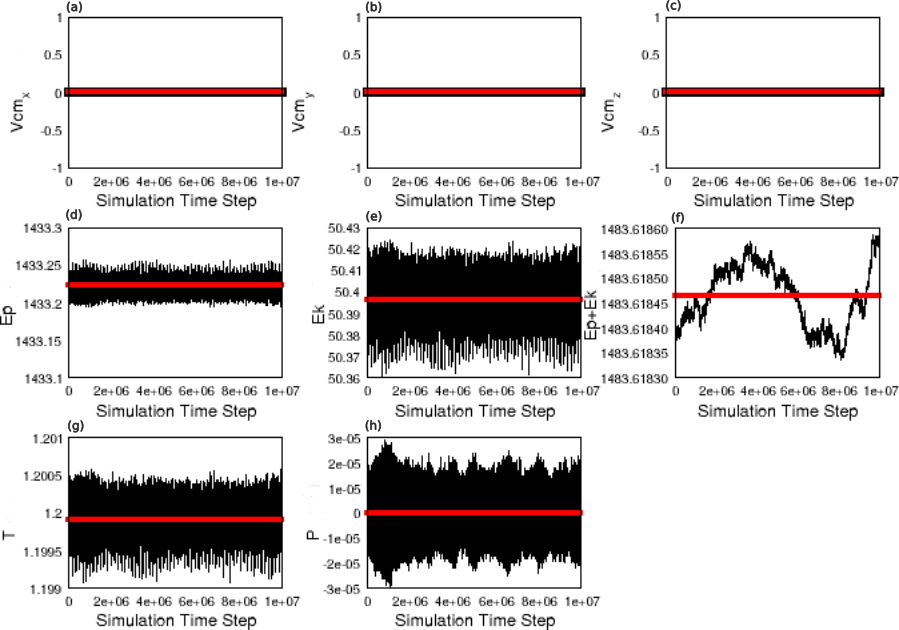}
 	\caption{Eight molecular dynamics simulation observables of a nanoparticle of 28 particles at equilibrium as a function of simulation time step in the NVE ensemble.}
 	\label{fig:observables28p}%
 \end{figure}

\begin{figure}[H]
 	\centering
 	\includegraphics[width=13.cm]{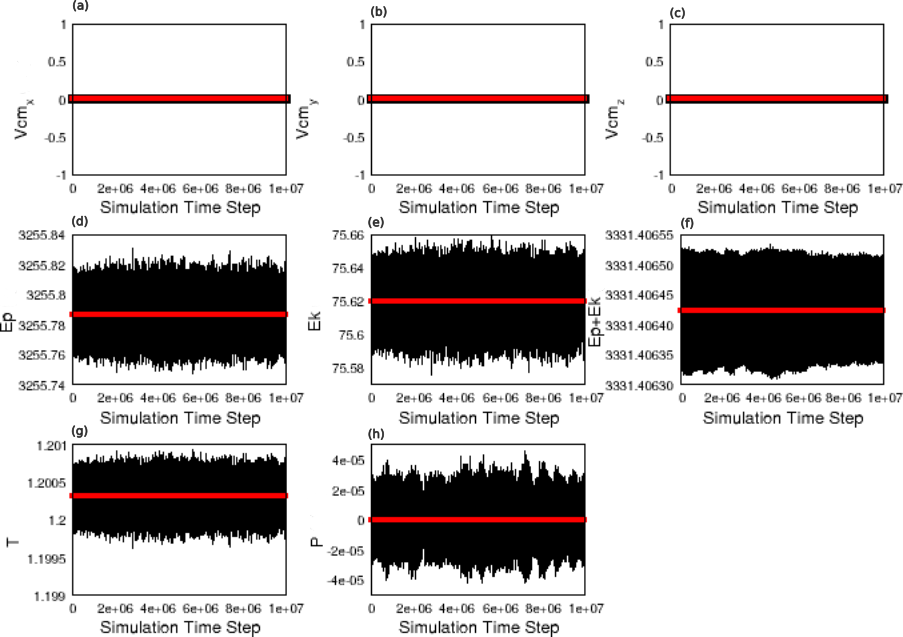}
 	\caption{Eight molecular dynamics simulation observables of a nanoparticle of 42 particles at equilibrium as a function of simulation time step in the NVE ensemble.}
 	\label{fig:observables42p}%
 \end{figure}

\begin{figure}[H]
 	\centering
 	\includegraphics [width=13.cm]{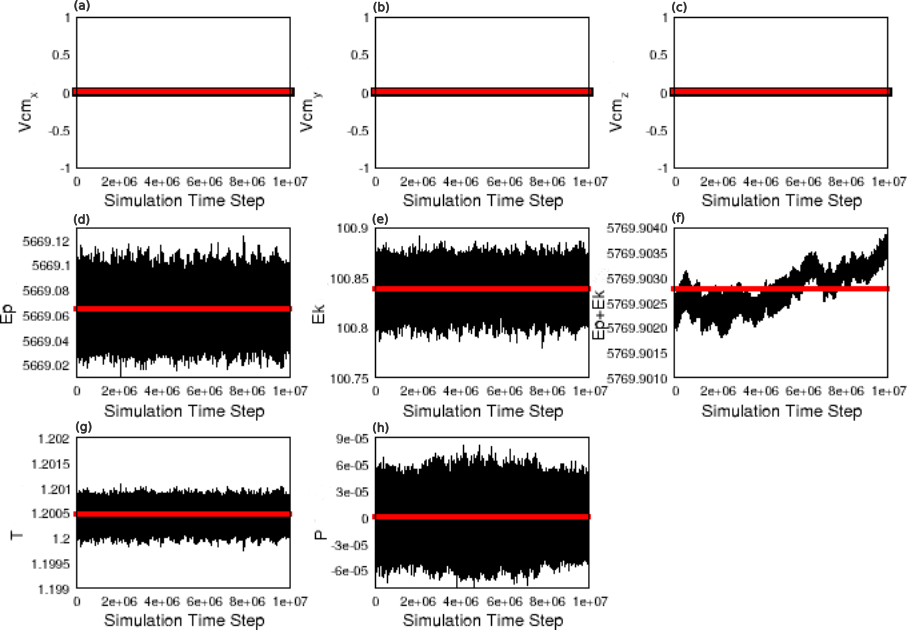}
 	\caption{Eight molecular dynamics simulation observables of a nanoparticle of 28 particles at equilibrium as a function of simulation time step in the NVE ensemble.}
 	\label{fig:observables56p}%
 \end{figure}

    Figure \ref{fgr:nanoparenergy} shows how the instantaneous total potential energy, $ \rm E_{p}$, the instantaneous total kinetic energy, $\rm E_{k} $, and the instantaneous total internal energy, $\rm E_{p}+E_{k} $, of the nanoparticles change with the total number of particles in a nanoparticle. When we look at the figure, we see that $\rm E_{k} $ increases linearly with the total number of particles in a nanoparticle. However, $\rm E_{p}+E_{k} $ and $ \rm E_{p}$ increases nonlinearly with the total number of particles in a nanoparticle.
 In Figures \ref{fig:observables28p}, \ref{fig:observables42p}, and \ref{fig:observables56p} the first three graph at the first line (a), (b), and (c) shows the x, y, and z-components of the instantantenous center of mass velocity of the nanoparticle of 28, 42, and, 56 atoms, $\rm vcm_{x}, \rm vcm_{y},\rm vcm_{z} $ as a function of the simulation time step, respectively. Since there is no external force, the instantaneous total momentum of the each system of three different nanoparticle is conserved upto the machine precision. In Figures \ref{fig:observables28p}, \ref{fig:observables42p}, and \ref{fig:observables56p}, (d), (e), and (f) shows the instantantenous total potential energy, $ \rm Ep $, the instantantenous total kinetic energy, $ \rm Ek $, and the instantantenous total internal energy, $ \rm Ep+Ek $,  of the three different nanoparticle as a function of the simulation time step, respectively. In Figures \ref{fig:observables28p}, \ref{fig:observables42p}, and \ref{fig:observables56p}, (g) and (h), instantaneous temperature, T, and instantaneous pressure, P, of the nanoparticles as a function of the simulation time step are depicted  respectively.
 
\begin{figure}[H]
 	\centering
 	\includegraphics[width=15cm]{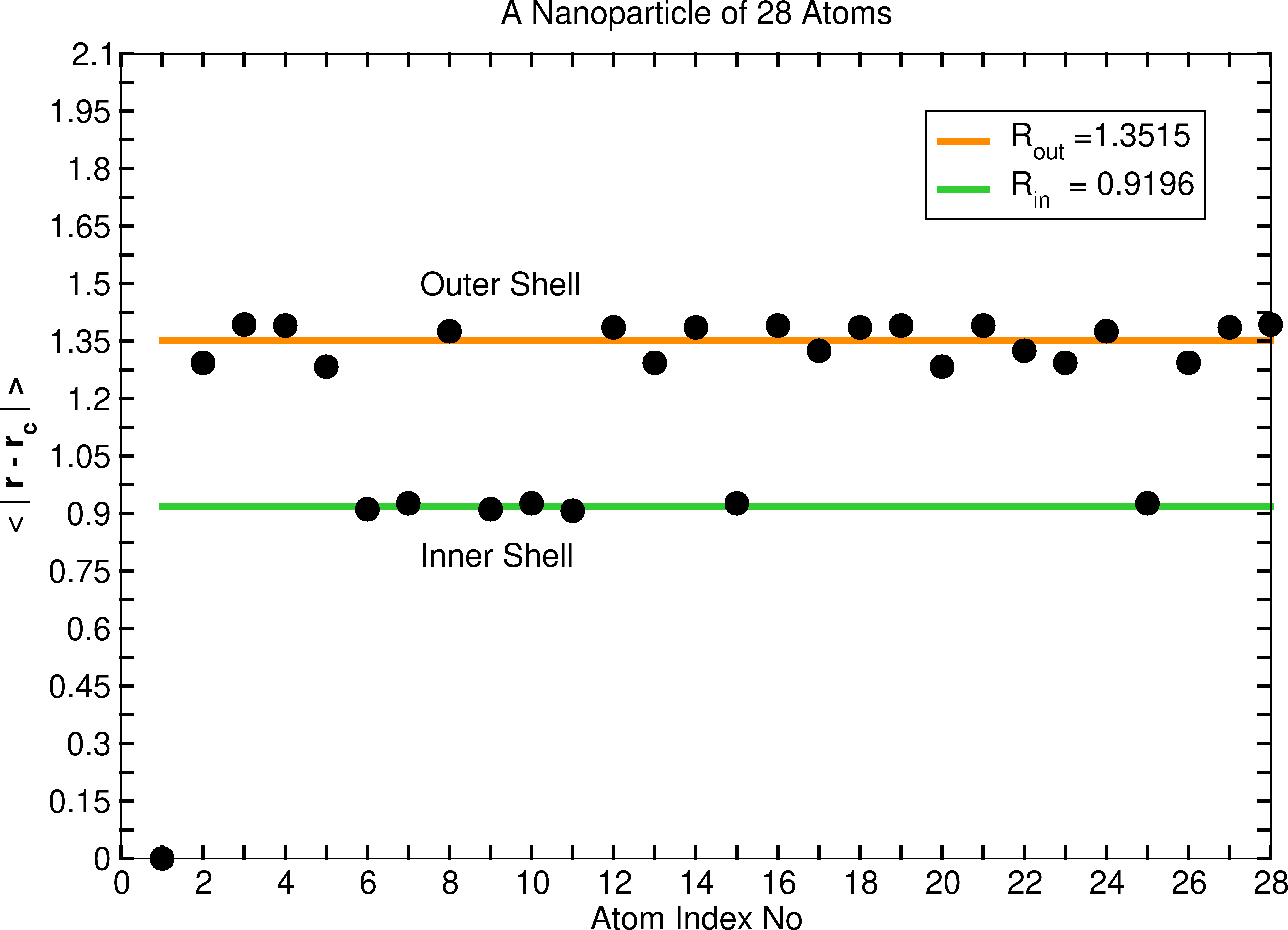}
 	\caption{The frame-averaged distance between the central atom and the other atoms belonging to same the nanoparticle of 28 particles as a function of atom index no. Green line and red lines represents $\rm R_{in} $ radius of the inner sphere and $\rm R_{out} $ radius of the outer sphere of two concentric shells of the same nanoparticle, respectively.}
 	\label{fig:RR28}%
 \end{figure}

\begin{figure}[H]
 	\centering
 	\includegraphics[width=15cm]{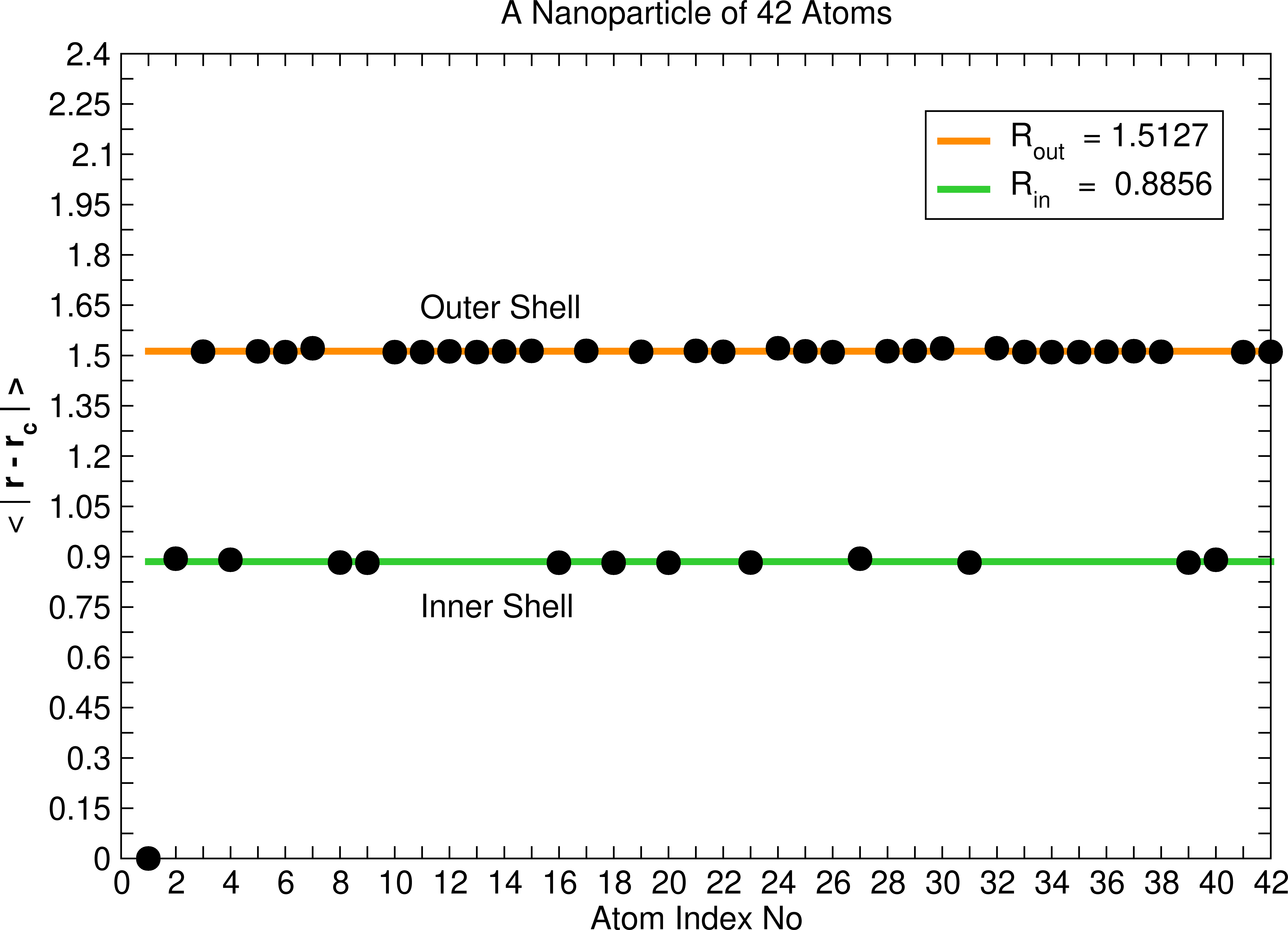}
 	\caption{The frame-averaged distance between the central atom and the other atoms belonging to the nanoparticle of 42 particles as a function of atom index no. Green line and red lines represents $\rm R_{in} $ radius of the inner sphere and $\rm R_{out} $ radius of the outer sphere of two concentric shells of the same nanoparticle, respectively.}
 	\label{fig:RR42}%
 \end{figure}

\begin{figure}[H]
 	\centering
 	\includegraphics[width=15cm]{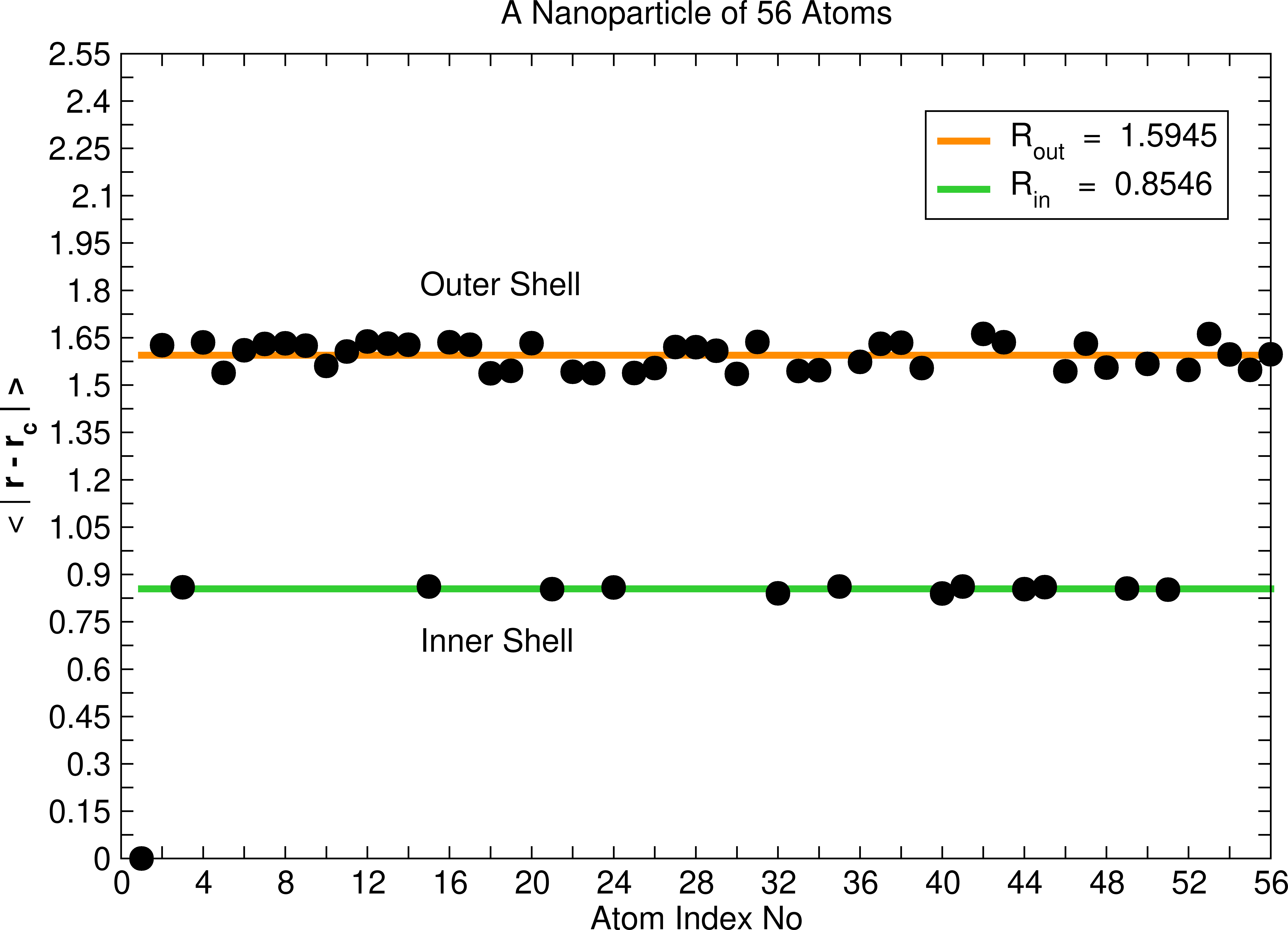}
 	\caption{The frame-averaged distance between the central atom and the other atoms belonging to the nanoparticle of 56 particles as a function of atom index no. Green line and red lines represents $\rm R_{in} $ radius of the inner sphere and $\rm R_{out} $ radius of the outer sphere of two concentric shells of the same nanoparticle, respectively.}
 	\label{fig:RR56}%
 \end{figure}

\begin{figure}[H]
	\centering
\includegraphics [width=11.cm]{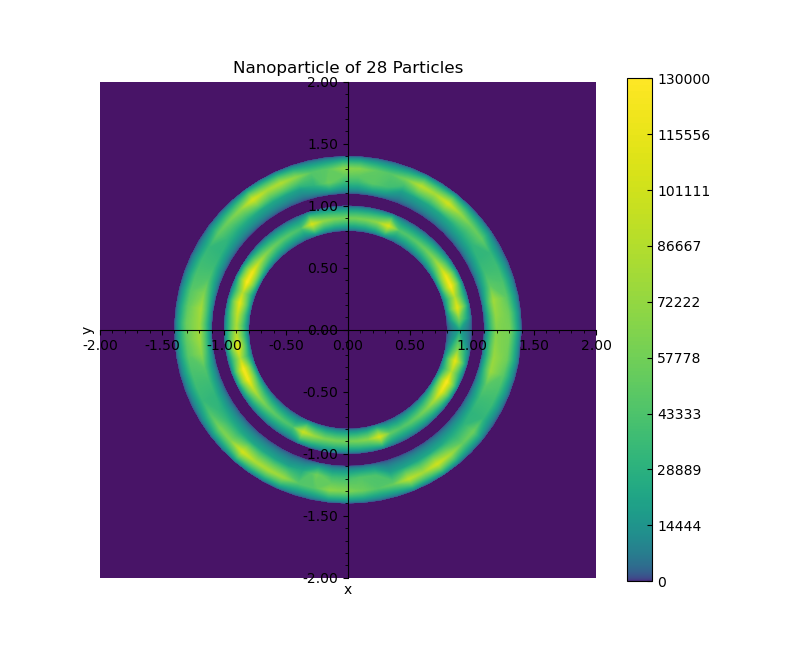}
\caption{Three dimensional angle dependent radial distribution function of a nanoparticle with 28 atoms. }
\label{fig:28pRDF}%
\end{figure}

\begin{figure}[H]
	\centering
	\includegraphics[width=9.cm]{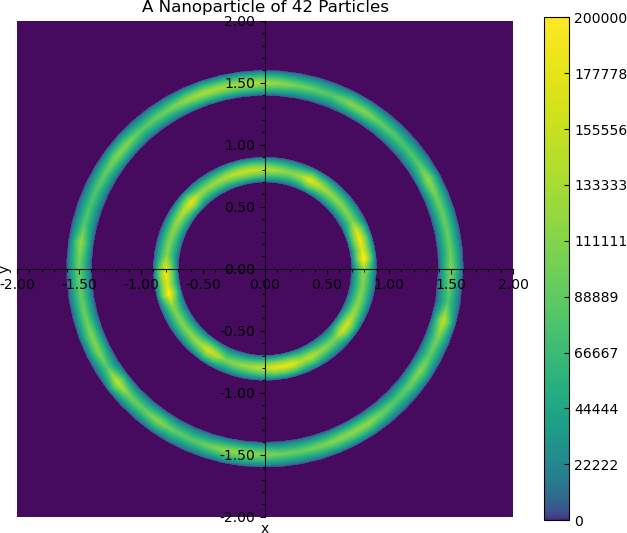} 
	\caption{Three dimensional angle dependent radial distribution function of a nanoparticle with 42 atoms. }
	\label{fig:42pRDF}%
\end{figure}

\begin{figure}[H]
	\centering
	\includegraphics[width=9.cm]{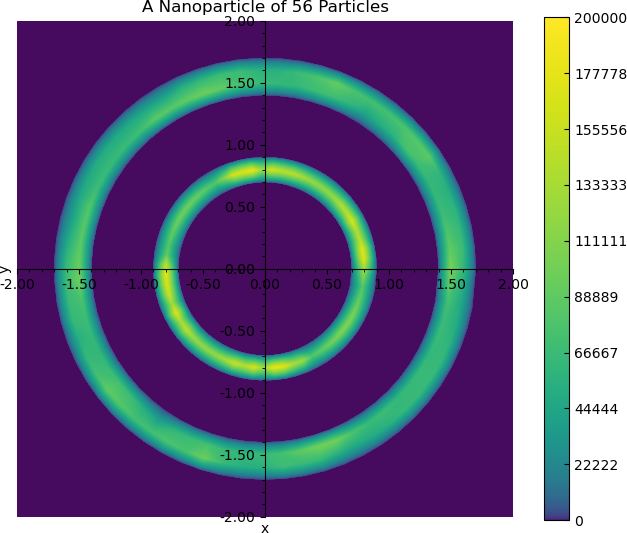} 
	\caption{Three dimensional angle dependent radial distribution function of a nanoparticle with 56 atoms. }
	\label{fig:56pRDF}%
\end{figure}

 We measured $\rm R_{in} $ and $\rm R_{out} $, the average inner and outer radius of two concentric shells of a nanoparticle with 28, 42, and 56 atoms, respectively. First of all, we performed a production simulation run of $\rm 1 \times 10^{6} $ time steps for the desired temperature $ \rm \frac{k_{B}}{\epsilon}T = 1.2$ for the system of only one nanoparticle in the center of the simulation box. In the course of this run, i.e., on the fly manner,  for every frame, the three dimensional separation vector between the central atom and one of the other atoms belonging to the same nanoparticle for three different nanoparticles, $ \vec{\rm r}-\vec{\rm r}_{c} $, are computed at each frame, where $ \vec{\rm r}_{\rm c}$ and $\vec{\rm r}$ is the position of the central atom and the other atom, which are measured relative to the origin of the simulation box. After this stage, we should apply the minimum image convention. And then, the calculated distances, $ |\vec{\rm r}-\vec{\rm r}_{c} |$, are assigned to the one dimensional array of 28, 42, and 56 elements in total, and then, each value in these arrays are summed with the values of the  previous run at each frame. After that, at the end of the simulation, we take the arithmetic mean of the last values in these arrays over the total frame number and in the end, we write these averaged values in a file. The Figures \ref{fig:RR28}, \ref{fig:RR42}, and \ref{fig:RR56} shows these frame-averaged distances, $ |\vec{\rm r}-\vec{\rm r}_{c}| $, as a function of atom index no. By using this graphs, we can calculate the average inner radius of the inner spherical shell, $\rm R_{in} $, and the average radius of the outer spherical shell, $\rm R_{out} $. When we look at the Figure \ref{fig:RR28}, we see that for the nanoparticle of 28 atoms,  there are seven atoms in the first inner spherical shell and twenty atoms in the second outer spherical shell. The distance between the shells is about 0.4319 units. The around radii of inner and outer shells are $ \rm R_{in}=0.9196 $ units and $ \rm R_{out}=1.3515$ units, respectively. The 42-atom nanoparticle in Figure \ref{fig:RR42} has twelve atoms in the first inner spherical shell and twenty-nine atoms in the second outer spherical shell. The shells are separated by around 0.6271 units. The approximate radii of inner and outer shells are $ \rm R_{in}=0.8856 $ units and $ \rm R_{out}=1.5127$ units, respectively. Twelve atoms make up the first inner spherical shell of the 56-atom nanoparticle in Figure \ref{fig:RR56}, while forty-three atoms make up the second outer spherical shell. The distance between the shells is approximately 0.7399 unit. The approximate radii of inner and outer shells are $ \rm R_{in}=0.8546 $ units and $ \rm R_{out}=1.5945$ units, respectively. When you look at the Figure \ref{fig:RR28} and the Figure \ref{fig:RR56}, the thicknesses of the outer spherical shells are more coarse than the thicknesses of the inner shells. We think that this thickness difference results from  the total particle number in a nanoparticle, intramolecular interactions such as FENE, and a truncated and shifted 12-6 Lennard-Jones potential energy function between the same nanoparticle atoms. In FENE-type interactions for three different nanoparticles, of course, we use three different $ \rm R_{0}$ and the same spring costant, $ \rm (\sigma^{2}/\epsilon) k=30$. These FENE radii should change the atom locations as well. In addition, the system temperature and system pressure can also affect this thickness distinction. If we think of these four different parameters: Nanoparticle atom number, FENE radius and FENE spring constant, temperature, and pressure, there should be a combined effect of them on the occurrence of different shell thicknesses, total atom numbers in outer and inner shells, and the spherical structures of nanoparticles consisting of two concentric spherical shells. These multiple factors acting together on discovering stable spherical nanoparticles with two concentric shells can be investigated in a much more detailed manner in a future paper.
 \\ \\
 For three distinct nanoparticles with 28, 42, and 56 atoms with two concentric shells, the Table \refeq{fig:tab1} displays the predefined minimum distance between any two atoms inside a nanoparticle, $\rm d_{0} $, the predefined radial distance from the central atom, $\rm r_{0}$, the maximum covalent bond length between the atom pair in the same nanoparticle, $\rm R_{0} $, the mean radius of the inner sphere, $ \rm R_{in}$, the surface area of the inner sphere, $ \rm A_{in}$, the mean radius of the outer sphere,  $\rm R_{out}$, and the surface area of the outer sphere, $ \rm A_{out}$. When we look at the Table \ref{fig:tab1}, $ \rm R_{in}$ and $ \rm A_{in}$ decreases linearly with the total atom number in a nanoparticle, while $ \rm R_{out}$ and $ \rm A_{out}$ increases nonlinearly with it.
 \begin{center}
 	\captionof{table}{Table shows the minimum distance between any two atoms inside a nanoparticle, $\rm d_{0} $, the radial distance from the central atom, $\rm r_{0}$, the maximum covalent bond length between the atom pair in the same nanoparticle for FENE potential energy function, $\rm R_{0} $, radius of the inner sphere, $ \rm R_{in}$, the surface area of the inner sphere, $ \rm A_{in}$, radius of the outer sphere, $ \rm R_{out}$, the surface area of the outer sphere, $ \rm A_{out}$ for a three different nanoparticles with 28, 42, and 56 atoms with two concentric shells.}
 	\begin{tabular}[H]{ |p{5cm}|p{1cm}|p{1cm}|p{1cm}|p{1cm}|p{1cm} |p{1cm}|p{1cm}|}   
 		\hline
 		\centering \vfill The Total Number of Particles in a Nanoparticle&\vfill
 		\centering $ \rm d_{0}$& \vfill
 		\centering $ \rm r_{0}$ & \vfill  
 		\centering $ \rm R_{0} $ &\vfill
 		\centering $ \rm R_{in}$ & \vfill
 		\centering $ \rm A_{in}$ & \vfill
 		\centering $ \rm R_{out}$ & \vfill 
 		\centering $ \rm A_{out}$  \cr 
 		\hline
 		\centering28 & \centering 0.55 &\centering 1.50 & \centering 1.52 & \centering $ \rm 0.9196 $ & \centering $ \rm 11 $ &\centering $ \rm 1.3515 $ & \centering 23 \cr
 		\hline
 		\centering42 & \centering 0.55 &\centering 1.60  & \centering 1.62& \centering $ \rm 0.8856 $ &\centering $ \rm 10  $ & \centering $ \rm 1.5127  $ & \centering 29 \cr
 		\hline
 		\centering56 & \centering 0.55 &\centering 1.69 & \centering 1.71 & \centering $ \rm  0.8546 $ &\centering $ \rm 9  $ & \centering $ \rm  1.5945  $ & \centering  32  \cr
 		\hline
 		
 	\end{tabular}
 	\label{fig:tab1}%
 \end{center} 
 We also calculated the angle dependent radial distribution functions (ARDF)  $\rm g(r, \phi, \theta) $ for three different nanoparticles by using a homemade Fortran code. The algorithm to create an angle dependent radial distribution function (ARDF)  $\rm g(r, \phi,\theta) $ of each nanoparticle in a simulation box without a polymer matrix is stated in the following:
				\begin{enumerate}
			\item Assign the value of the radial distance mesh resolution, $ \rm \Delta rad$,\:
			\item Assign the value of the azimuthal angle mesh resolution, $ \rm \Delta theta$,\:
			\item Assign the value of the polar angle mesh resolution, $ \rm \Delta phi$,\:
			\item Define the total mesh number in the direction of the radial unit vector, 
			$\rm nrad=\frac{\rm L}{\rm 2\Delta rad}  $,  \:
			\item Define the total mesh number in the direction of the azimuthal angle unit vector, 
			$\rm ntheta=\frac{\rm2\pi }{\rm \Delta theta}  $,  \:
			\item Define the total mesh number in the direction of the polar unit vector, 
			$\rm nphi=\frac{\rm \pi }{\rm \Delta phi} $,  \:
			\item Find the mesh index no, corresponding to the great circle at the polar angle $ \rm \phi=90^{\circ}$, $\rm gcircle$, as you can see in the Figure \ref{fig:gcircle}.\:
				\begin{figure}[H]
				\centering
		    	\includegraphics[height=6cm]{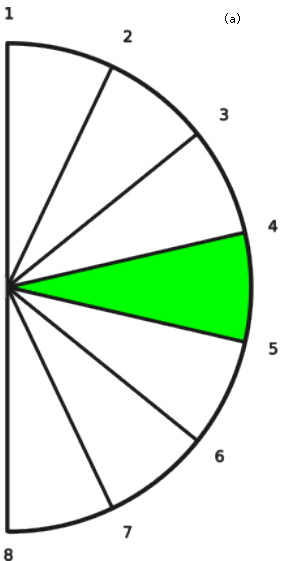}
				 \includegraphics[height=6cm]{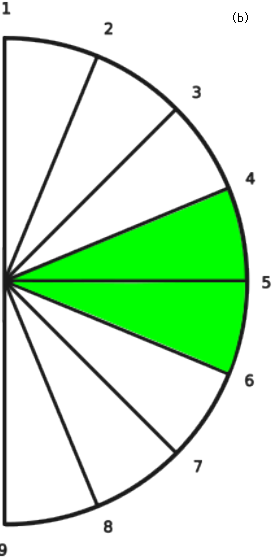}
				\caption{if nphi is even, then $ \rm gcircle= \frac{nphi}{2} $, as seen in the left-hand side Figure  (a), and	if nphi is odd, then $ \rm gcircle= \frac{nphi+1}{2} $, as seen in the right-hand side Figure (b).}
				\label{fig:gcircle}%
			\end{figure} 
			if nphi is odd,  then $ \rm (gcircle-2)\Delta phi< \phi < gcircle \Delta phi$, \\
			if nphi is even, then $ \rm (gcircle-1)\Delta phi< \phi < gcircle \Delta phi$
			\item Allocate an array with three dimensions of three indices, $ \rm g(0:nframe, 1:nrad, 1:ntheta )$,
			\item A loop from 1 to nframe for nf index, \:
			\item Read atom positions in 3D,
			\item A loop for the nanoparticle atoms from 1 to Npar for Np index, \:
			\item For each nanoparticle atom, find new 3D separation vector components relative to the central atom of the nanoparticle, $\vec{\rm r}-\vec{\rm r}_{c}$.
			\item Use minimum image convention,\;
			\item Calculate the distance $\rm d=|\vec{\rm r}-\vec{\rm r}_{c}|$. If d is greater than the L/2 then go  to the step 11, if not, go to the step 15,\;
			\item Transform cartesian coordinates $ \rm (x,y,z) $ to spherical coordinates $ \rm (d,\phi,\theta) $:\\
			$ \rm d=\sqrt{x^2+y^2+z^2}$ \\ 
			$ \rm \phi=arccos(z/d)$ \\
			$ \rm \theta=arctan(y/x)$
			\item Find mrad, mphi, mtheta values, which are mesh (box) indices that $ \rm (d, \phi, \theta)$ point corresponds to,\;\\
			$\rm mrad=\frac{\rm d}{\rm \Delta rad}  $\\
			$\rm mphi=\frac{\rm \phi}{\rm \Delta phi}  $\\
			$\rm mtheta=\frac{\rm \theta}{\rm \Delta theta}  $
			\item If (nphi is odd number) and (mphi is equal to gcircle or mphi is equal to gcircle-1), then increase\\
			$ \rm g(nf, mrad, mtheta )=g(nf, mrad, mtheta )+1$\\
		    If (nphi is even number) and (mphi is equal to gcircle), then increase \\
			$ \rm g(nf, mrad, mtheta )=g(nf, mrad, mtheta )+1$\;
				\item If Np is equal to Npar, then go to Step 19, if not, go to step 11, 
		    \item If nf is equal to nframe, then 
		 \item[]  A1 loop from 1 to nrad for i index
		\item[] 	 A2 loop from 1 to ntheta for j index
		\item[]  sum=0
		\item[]  A3 loop from 1 to nframe for k index	             	\item[]     sum=sum+g(k, i, j)   	\item[]   end loop A1	\item[]   g(0, i, j)=sum 
		    	\item[]  end loop A2	\item[]  end loop A3
		 \item[] and go to 20, if nf is not equal to nframe, go to step 9
		\item Start to write (xx, yy) cartesian points corresponding to (nrad, ntheta) boxes in an azimuthal coordinate plane at $ \rm \phi=90^{\circ} $ and the relative number density, $\rm \frac{\rho_{mesh}}{\rho_{par}} $ , (relative means that it is normalized to the number density of the nanoparticle = $\rm \rho_{par}$= (Npar-1)/Volume, also we subtract one from Npar because we do not consider the central atom of the nanoparticle when the new coordinates are calculated with respect to it as stated at step 11) at the (xx, yy) point,
	\item[] B1 loop from 1 to nrad for r index
	 \item[] Define discrete differential volume element:
	 \begin{eqnarray}
	\rm  \text{	if(nphi is odd),} \;\; \Delta V=	\int_{t \Delta theta}^{t \Delta theta+\Delta theta}d\theta \, \int_{(gcircle-2)\Delta phi}^{gcircle \Delta phi} \sin \phi d\phi \, \int_{r \Delta rad- \Delta rad }^{r \Delta rad} R^2  dR \,\\
	 \text{then:}\;\; \rm \Delta V= \frac{1}{3}(r^{3}-(r-1)^{3})
	 (\Delta rad )^{3} (cos((gcircle-2)\Delta phi)-cos(gcircle \Delta phi)) \Delta theta \\
	 	\rm  \text{	if(nphi is even),} \;\; \Delta V=	\int_{t \Delta theta}^{t \Delta theta+\Delta theta}d\theta \, \int_{(gcircle-1)\Delta phi}^{gcircle \Delta phi} \sin \phi d\phi \, \int_{r \Delta rad- \Delta rad }^{r \Delta rad} R^2  dR \,\\
	 \text{then:}\;\; \rm \Delta V= \frac{1}{3}(r^{3}-(r-1)^{3})
	 (\Delta rad )^{3} (cos((gcircle-1)\Delta phi)-cos(gcircle \Delta phi)) \Delta theta
	 \end{eqnarray}
	    
\item[]B2 loop from 1 to ntheta for t index
\item[]	if(nphi is odd), then
\item[]	$ \rm xx=(r-1)\Delta rad \;sin((gcircle-1)\Delta phi) \;cos((t-1)\Delta theta)$ 
\item[]	$ \rm yy=(r-1)\Delta rad \;sin((gcircle-1)\Delta phi) \;sin((t-1)\Delta theta)$      
\item[]	if(nphi is even), then 
\item[]	$ \rm xx=(r-1)\Delta rad \;sin((gcircle-1+0.5)\Delta phi) \;cos((t-1)\Delta theta)$ 
\item[]	$ \rm yy=(r-1)\Delta rad \;sin((gcircle-1+0.5)\Delta phi) \;sin((t-1)\Delta theta)$  
\item[]if t is equal to ntheta and nphi is odd, then
\item[]	$ \rm xx=(r-1)\Delta rad \;sin((gcircle-1)\Delta phi) \;cos((1-1)\Delta theta)$ 
\item[]	$ \rm yy=(r-1)\Delta rad \;sin((gcircle-1)\Delta phi) \;sin((1-1)\Delta theta)$ 
\item[]if t is equal to ntheta and nphi is even, then
\item[]	$ \rm xx=(r-1)\Delta rad \;sin((gcircle-1+0.5)\Delta phi) \;cos((1-1)\Delta theta)$ 
\item[]	$ \rm yy=(r-1)\Delta rad \;sin((gcircle-1+0.5)\Delta phi) \;sin((1-1)\Delta theta)$  
\item[]if t is equal to ntheta, then write xx, yy, 
 $\rm \frac{g(0,r,1)}{\Delta V \rho_{ par}} $ to a file
 \item[]if t is not equal to ntheta, then write xx, yy, 
 $\rm \frac{g(0,r,t)}{\Delta V \rho_{ par}} $ to the file
\item[] end loop B1
\item[] end loop B2
\end{enumerate}
By using the just above algorithm, we calculate the $\rm g(r,\phi,\theta)$ for three different nanoparticles of 28, 42, 56 particles. The Figures \ref{fig:28pRDF}-\ref{fig:56pRDF} show three different 2D contour maps corresponding to these nanoparticles. In these graphs the origin corresponds to center of the central atom belonging to the relevant nanoparticle. All of the contour maps include two concentric spherical shells. For three different contour map we use the mesh resolution of 0.1 units for the radial, polar, and azimuthal distances. The contour maps were calculated for 3000000 simulation time steps. As shown by three figures, the outer shells for the nanoparticles of 28 and 56 atoms are thicker than the outer shell of the nanoparticle of 42 atoms. Also the average radii of the outer and inner shells of the nanoparticles matches the values found by the Figures \ref{fig:RR28}-\ref{fig:RR56}. In addition, all of the outer shells' contour boundaries are compatible with the FENE radii, $\rm R_{0}=1.52, \;1.62, \;1.71$ for the nanoparticles of 28, 42, and 56 atoms, respectively. Also we can take these three $\rm R_{0} $ as the average radii of three nanoparticles. We conclude that by using FENE potential energy function with different FENE radii and the same FENE spring constant, $ \rm (\sigma^{2}/\epsilon) k=30$, one can construct spherical nanoparticles with two concentric hollow shells.    
\subsection{Monomer-Nanoparticle Interactions}
An atom-pair between a nanoparticle atom and a monomer also interacts with a \textit{modified}, truncated, and shifted 12-6 Lennard-Jones potential energy function \cite{Barrat1999} as well (See Figure \ref{fgr:interactions} and Eq.\ref{eq:UmCnpa}-\ref{eq:ModfULJ}). $ \rm C_{\alpha \beta}$ rules the attractive part of the potential $\rm U_{\rm \alpha \beta} (\rm r) $, therefore the larger the $ \rm C_{mn}$ is, the stronger the affinity between the nanoparticles and monomers are. By increasing or decreasing the $ \rm C_{mn}$, we can obtain depletion layers without polymers and polymeric adsorption layers  with different infinitesimal thicknesses formed on the nanoparticle surface and extends into the polymeric dispersion medium.    
\begin{figure}[H]
	\centering
\includegraphics[height=7cm]{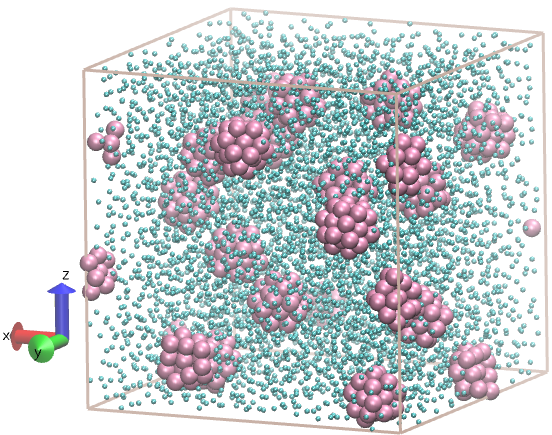}
	\caption{Molecular dynamics simulation snapshot of a bulk nanofluid of 6160 particles at equilibrium. The system composed of 20 nanoparticles (pink) inside the base polymer melt of 5600 monomers (cyan). The approximate volume of the cubic system is equal to $\rm V=7558$. Thanks to the periodic boundary conditions, we can see the atom groups from one atom to 28 atoms  belonging to the same nanoparticle on the different sides of the simulation box (light brown) in the snapshot. For visualization aim, we use VMD \cite{Humphrey1996}.}
	\label{fgr:nanofluid}
\end{figure}
\noindent
\subsection{Combining Polymers and Nanoparticles}
At this point, we mix the nanoparticles with the polymer matrix (see Figure \ref{fgr:nanofluid}). A homemade Fortran code is written for this purpose. The following is the algorithm for this:

\begin{enumerate}
	\item Define the total monomer number in a polymer chain, $\rm N_{mon}$, total polymer melt atom number (total liquid atom number=total monomer number) in the nanofluid, $\rm N_{liquid}$, the total atom number in a nanoparticle, $\rm N_{p}$, the total number of nanoparticles, $\rm N_{par}$, total atom number (liquid and nanoparticle atoms) in the nanofluid, $\rm N_{atom}$,
	\item Define the number density of the nanofluid, $\rm \rho_{nano}$, 
	\item Define the around radius af the nanoparticle, $\rm r_{p}$
	\item Define the side length of the cubic volume occupied by a single atom in the nanofluid (i.e., particle diameter) by calculating $\rm a_{1}=  (1/\rho_{nano})^{1/3}$
	\item Define the around distance between two atoms in the y direction or z direction, $\rm a_{2}=2r_{p}+a_{1}$
	\item Evaluate the number of monomers in the x, y, and z directions using the formula $\rm n_{a}=(N_{liquid})^{1/3}$,
	\item Determine the total number of polymer chains in the x direction, $\rm n_{p_{x}}= n_{a}/N_{mon} $, ($\rm  n_{p_{x}}$ must be greater than or equal to 2), 
	\item Determine the total number of monomers in the x, y, and z directions, respectively, $ \rm n_{x}=n_{p_{x}} N_{mon} $, $ \rm n_{y}=n_{a} $, $ \rm n_{z}=n_{a}$,
	\item Evaluate the total number of nanoparticles, $ \rm N_{p_{t}}=(n_{p_{x}}-1)n_{y}n_{z}$, 
	\item Determine the total number of nanoparticle atoms, $\rm N_{par_{1}}=N_{p_{t}}N_{p}$,
	\item Determine the total liquid atoms, $\rm N_{liquid{1}}=n_{x}n_{y}n_{z} $,
	\item Determine the total number of atoms, $ \rm N_{atom_{1}}=N_{par_{1}}+N_{liquid{1}}$, 
	\item Calculate the distance between the first and the last atom in the x direction by using the formula 
		\item[] $ \rm L_{x}=(n_{p_{x}}(N_{mon}-1)a_{1})+
	((n_{p_{x}}-1)2 r_{p} 2 a_{1})$,
	\item Calculate the distances between the first and the last atom in the y direction and in the z direction by using the formulas respectively, 
	\item[]$ \rm L_{y}=(n_{y}-1)a_{2}$
	\item[]$ \rm L_{z}=(n_{z}-1)a_{2}$
	\item Define the simulation box sizes and volume as follows
	\item[] $\rm box(x)=L_{x}+a_{1}$
	\item[] $\rm box(y)=L_{y}+a_{2}$
	\item[] $\rm box(z)=L_{z}+a_{2}$
\item Find the largest value, maxval, of box(x), box(y), and box(z),
\item Assign the maxval to box(1), box(2), and box(3), and calculate the simulation box volume, Volume, respectively,
\item[] $\rm box(x)=maxval$
\item[] $\rm box(y)=maxval$
\item[] $\rm box(z)=maxval$
\item[] $\rm Volume=box(x)box(y)box(z)$	
\item Allocate 3D coordinates of the nanofluid, r($ \rm N_{atom_{1}}$,3), 3D coordinates of the equilibrated nanoparticle, $\rm r_{p}=(\rm N_{p},3)$, and residue numbers of the atoms, residue($ \rm N_{atom_{1}}$),   
\item Read the equilibrated nanoparticle coordinates in three dimensions. 
\item[] If the center of mass coordinates of the nanoparticle is not at the origin (0,0,0), we should bring the center of mass coordinates to the origin. If we don't, before stowing the nanoparticles between the polymer chains, we have to subtract the nonzero center of mass of the equilibrated nanoparticle from these coordinates on which we will put the nanoparticle's center of mass. Right after doing this, we have to translate the every atom of the nanoparticles to this new position by \textit{this difference}. We have to do this operation for each nanoparticle. If we do not want to do this and the center of mass coordinates of the nanoparticle is not equal to zero, we have to set it to (0,0,0) right after reading the coordinates of the equilibrated nanoparticle:
\item Find the center of mass of the coordinates of the nanoparticle and check if they are equal to zero or not, if not, calculate the new position components of the nanoparticles with respect to the observer at the nanoparticle's nonzero center of mass.
\item $\rm k2=N_{liquid_{1}}$, which is a counter for indexing the nanoparticle atoms,
\item[]A1 loop from 1 to $\rm n_{z}$ for z index
\item[]A2 loop from 1 to $\rm n_{y}$ for y index
\item[]$\rm c_{x}=0$, which is a counter for the  nanoparticle number
\item[]A3 loop from 1 to $\rm n_{x}$ for x index
\item[]k=k+1, which is a counter for total atom number (liquid and nanoparticle atoms),
\item[]m=m+1, which is a counter for liquid atom numbers,
\item[]if($\rm x \le N_{mon}$)
$\rm r(m,x)=(-L_{x}*0.5)+((x-1)a_{1})$,
\item[]if($\rm mod(x,N_{mon})=0 $ and  x is not equal to $\rm n_{x}$)then
\item[]$\rm c_{x}=c_{x}+1 $
\item[] $\rm x_{shift}=(-L_{x}*0.5)+((x-c_{x})a_{1})+(r_{p}+a_{1})+((c_{x}-1)(2 r_{p}+ 2 a_{1}))$
\item[] $\rm y_{shift}=(-L_{y}*0.5)+((y-1)a_{2})$
\item[] $\rm z_{shift}=(-L_{z}*0.5)+((z-1)a_{2})$
\item[]A4 loop from 1 to $\rm N_{p}$ for i index
\item[]k=k+1, 
\item[]$\rm k_{2}=k_{2}+1$
\item[] $\rm r(k_{2},x)=r_{p}(i,x)+x_{shift} $
\item[] $\rm r(k_{2},y)=r_{p}(i,y)+y_{shift} $
\item[] $\rm r(k_{2},z)=r_{p}(i,z)+z_{shift} $
\item[] End Loop A4
\item[] End if
\item[]if($\rm x > N_{mon} $) then
\item[]if($\rm mod(x,N_{mon})=1 $) $ \rm bb=\lfloor x/N_{mon} \rfloor$
\item[]$\rm r(m,x)=(-L_{x}*0.5)+(((x-1)-bb)a_{1})+(bb(2 r_{p}+ 2 a_{1}))$
\item[] endif 
\item[] $\rm r(m,y)=(-L_{y}*0.5)+((y-1)a_{2})$,
\item[] $\rm r(m,z)=(-L_{z}*0.5)+((z-1)a_{2})$,
\item [] if (k is equal to $ \rm N_{atom_{1}}$) go to 20,
\item[] End Loop A1
\item[] End Loop A2
\item[] End Loop A3
\item  Check the center of mass position of the nanofluid, it should be zero up to the machine precision.
\item  Apply periodic boundary conditions,
\item  Check the center of mass position of the nanofluid again, it shouldn't change and still be zero up to the machine precision.
\item By using an uniform random number generator, \textbf{ran1} \cite{Press2000}, select $ \rm N_{par}$ atoms randomly by using the residue numbers of the nanoparticles:
\item[]idum=-1
\item[]resleft=residue($ \rm N_{liquid_{1}}+1$)
\item[]resright=residue($ \rm N_{atom_{1}}$)
\item[]Allocate the array, $\rm p_{r}( N_{par}) $, containing  $\rm N_{par} $ residue indexes in total, 
\item[]A5 Loop from 1 to $\rm N_{par} $ for index i
\item[]tmp=(\textbf{ran1(idum)}(resleft-resright))+resright
\item[]$\hookrightarrow$ which converts the number in $\rm (0,1)$ to the number in $\rm (residue(N_{liquid_{1}}+1),residue(N_{atom_{1}})) $ by using the Eq.\ref{eq:convert1} and Eq.\ref{eq:convert2},
\item[]$\rm p_{r}( N_{par})=tmp $
\item[]End Loop A5
\item[]Check whether or not the numbers appear more than once, if so, change the initial seed, idum, for \textbf{ran1} algorithm \cite{Press2000}:
\item[]A6 Loop from 1 to $\rm N_{par}-1 $ for index i
\item[]A7 Loop from i+1 to $\rm N_{par} $ for index j
\item[]if ($\rm p_{r}(i)= p_{r}(j)$) then 
 \item[]id0=id0-1
 \item[]idum=id0
 \item[]Go to Loop A5
 \item[]End if
 \item[]End Loop A6 
 \item[]End Loop A7 
 \item Write the 3D coordinates of the nanofluid into a file as you can see in Figure \ref{fig:mixture}.
\item[]
 \begin{figure}[H]
	\centering
	\subfloat[][\centering Perspective view of the nanofluid of 6480 monomers (blue) and 324 nanoparticles (pink) before extracting the monomers and the nanoparticles.] {{\includegraphics[width=8cm]{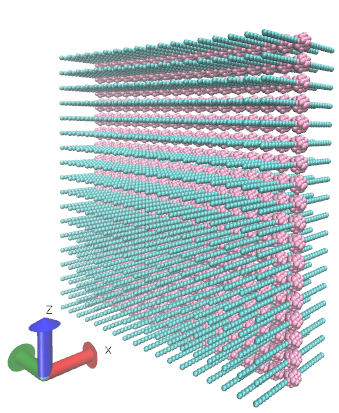} }} \hspace{0.001cm}
	\subfloat[][\centering Orthographic view of the nanofluid of 6160 monomers (blue) and 20 nanoparticles (pink) right after extracting the monomers and the nanoparticles.] {{\includegraphics[width=10cm]{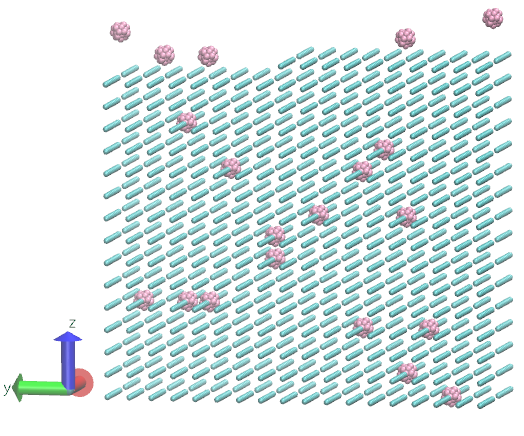} }}  \hspace{1cm}
	\caption{Visualization of the nanofluid before extracting the monomers and the nanoparticles (a) and right after removing the monomers and the nanoparticles (b). In (b), we subtract 320 monomers and we choose 20 nanoparticles randomly. For rendering purpose, we use VMD \cite{Humphrey1996}. }%
	\label{fig:mixture}%
\end{figure}
\noindent
 \end{enumerate}
\section{Radial Distribution Functions for Homogenous Systems}
A radial distribution function (pair correlation function, pair distribution function), g(r), gives information about internal structures of systems consisted of atoms.
Radial distribution function measures how atoms arrange themselves around one another, i.e., they give information about the local structure. While g(r) can be measured experimentally for atomic substances using neutron-scattering or X-ray techniques, it can be quickly and easily computed from molecular dynamics simulation trajectories, because, as we mentioned before, we can obtain the positions of individual particles as a function of time during a molecular dynamics simulation \cite{Haile1992}.
\begin{figure}[H]
	\centering
	\includegraphics[height=5cm]{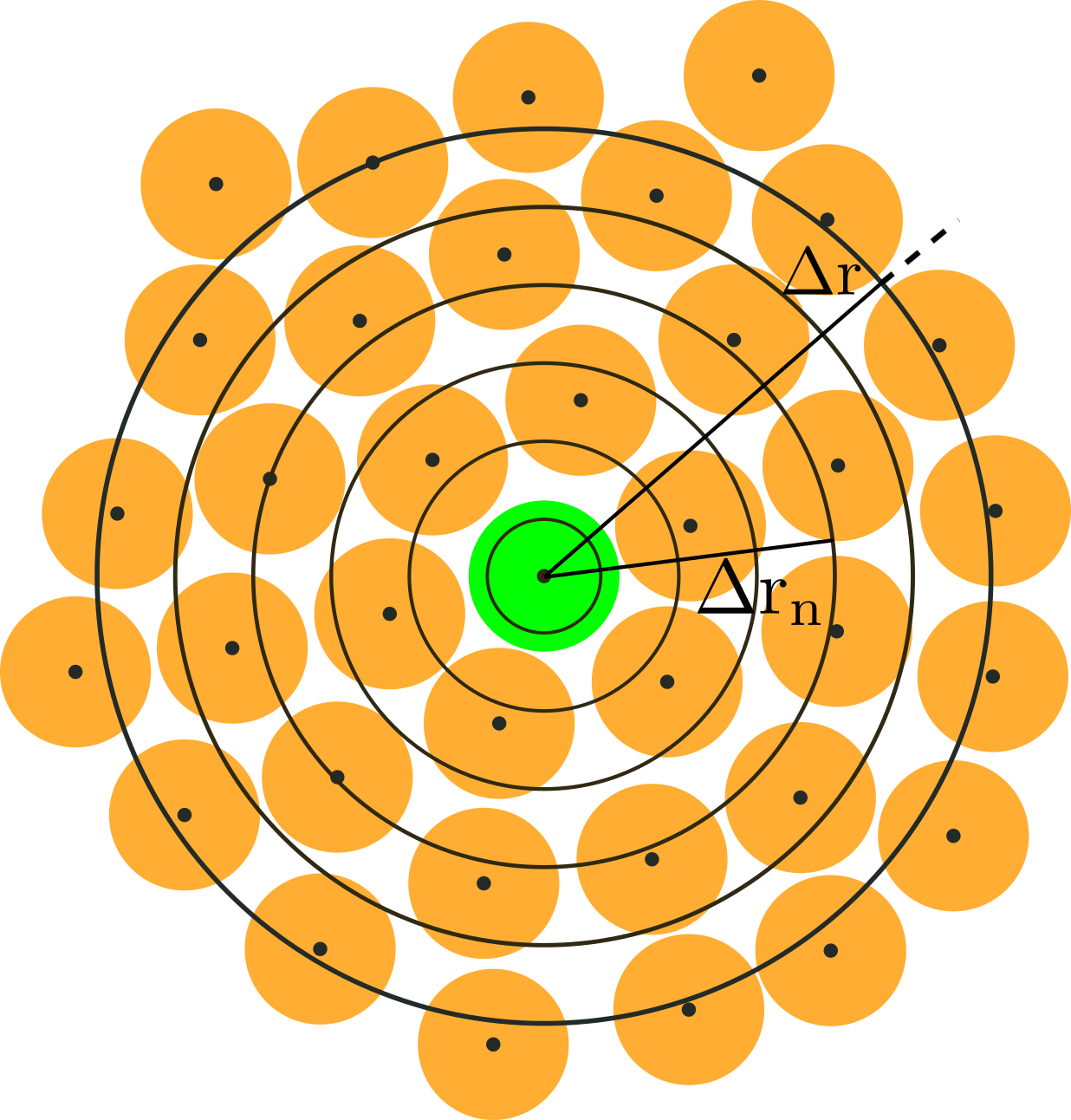}
	\caption{Schematic illustration of the first six  spherical coordination spheres (equidistant concentric circles (black)) for a central liquid state atom (green). The radial distance between spherical shell surfaces is equal to $\rm \Delta  r$. Radial distance from the central atom (origin) is equal to $\rm \Delta r_{n}= n \Delta r, \; n=1,2, ..., n_{max}, \; n_{max}=\frac{L/2}{\Delta r},$ L is the side length of the cubic simulation box.}
	\label{fgr:rdf1}
\end{figure}
\noindent
For molecular dynamics simulation of homogeneous isotropic liquid, we can define the following one dimensional local radial distribution function,
\begin{equation}
\rm g(\Delta r_{n})=\rm \frac{\langle\rho(\Delta r_{n})\rangle}{\rho_{b}}
\end{equation}
as a local number density distribution function for the  $ \rm n^{th}$ spherical shell, which is normalized to one by dividing the local average radial number density profile, $\rm\langle\rho(\Delta r_{n})\rangle $, which is a function of the distance $\rm \Delta r_{n}$, which is $\rm \Delta r_{n}= n \Delta r, \; n=1,2, ..., n_{max}, $ $\rm  n_{max}=\frac{L/2}{\Delta r}$, from the origin (i.e., the central atom) by the bulk number density, $\rm \rho_{b}$ (see Figure \ref{fgr:rdf1}). The ensemble average symbol $\langle \cdots \rangle $ denotes  averaging over total frame number and total atom number. We can define  $\rm\langle\rho(\Delta r_{n})\rangle $ as the following:
\begin{equation}
\rm \langle\rho(\Delta r_{n})\rangle = \rm \frac{N_{total}(\Delta r_{n} )}{V_{shell}(\Delta r_{n} )N_{liquid}N_{frame}},
\end{equation}
where $\rm N_{total}(\Delta r_{n} )$ is the total number of liquid atoms in the $ \rm n^{th}$ bin, which is found by adding the number of atoms in the  $ \rm n^{th} $ shell for each frame and each liquid at the center around which the spherical bins are defined, $  \rm V_{shell}(\Delta r_{n} )$, which is a function of $\rm \Delta r_{n}  $ too, 
\begin{equation}
\rm V_{shell}(\Delta r_{n} )=\rm \frac{4}{3}\pi(\Delta r_{n}^{3}-(\Delta r_{n}-\Delta r)^{3}) 
\end{equation}is the volume of the  $ \rm n^{th}$ spherical shell with thickness $\rm \Delta r $, $\rm N_{liquid}$ is the total number of liquid atoms in the simulation box, and $\rm N_{frame}$ is the total number of the simulation frames. Bulk number density for a target atom type is given by as follows:
\begin{equation}
\rm \rho_{b}=\rm \frac{N_{liquid}-1}{V_{sim}},
\label{eq:rho}
\end{equation}
where $\rm V_{sim} $ is the volume of the cubic simulation box.
In the numerator in Eq.\ref{eq:rho}, numeral one must be 
subtracted from $\rm N_{liquid} $ because we have to exclude the central reference atom while counting the bulk phase atoms.\\ \\
We prepared a test Lennard-Jones liquid system to investigate the static (in equilibrium) behavior of the Lennard-Jones liquid of 500 atoms. To reach an equilibrium state at a system temperature of $\rm T = 0.722$ and a number density of $\rm \rho= 0.8442$, i.e., the number of particles per unit simulation box volume, the velocity components of the particles are adjusted at certain simulation time-intervals according to the desired temperature value (velocity scaling). Using a microscopic interpretation of the temperature, we can adjust the velocities as follows: Absolute temperature in Kelvin, T, is proportional to the average translational kinetic energy of particles, K, in three dimensional cartesian space:
\begin{equation}
\rm 	\langle K \rangle =	\langle \frac{1}{2} m_{i}  {{\textbf{v}}_{i}}^{2} \rangle = 3N \frac{k_B T}{2}
\end{equation}
where N is the total atom number, Boltzmann constant $\rm k_{B}=1.38 \times 10^{-23} J/(K.mol)$, $\rm m_{i}$ is the $\rm i^{th}$ particle mass, and $\rm \mathbf{v_{i}}$ is the $\rm i^{th}$ particle velocity vector. The factor of 3 arises from the fact that a single-atomic molecule has only three different translational degrees of freedom. $\langle \; \rangle  $ represents the ensemble average. This microscopic definition of temperature is used to define the instantaneous time-dependent temperature of our system in three dimensions as follows: \begin{equation}
\rm	T(t) = \frac{2}{ k_{B }(3N-3)} \sum_{i=1}^{N} \frac{1}{2} m_i \mathbf v_i(t)^{2}
\end{equation}
where $\rm N$ is the total number of particles, 
$\rm \mathbf{v}_{i}(t)$ is the instantaneous velocity vector of the i. molecule. $ \rm T(t)$ is the system's instantaneous kinetic temperature, which is proportional to the instantaneous average translational kinetic energy per particle in our simulation box. $ \rm 3N-3$ is the number of degrees of freedom with zero total momentum \cite{frenkel2023}. Using this definition of instantaneous temperature, the velocities calculated during the simulation are scaled as follows:
\begin{equation}
\rm	v_{ij}^{'}(t) = \sqrt{\frac{\rm T}{\rm T(t)}}\; v_{ij}(t), \quad i = 1, \dots, N, \quad j = x, y, z
	\label{eq:scaled}
\end{equation}
where the index i represents the atom number, while the index j represents the spatial dimension. $\rm T$ is the desired system temperature (as a parameter), $\rm v_{ij}^{'}(t) $ is the $\rm  j^{th} $ component of the scaled instantaneous velocity vector of the $\rm i^{th}$ particle, $\rm T(t)$ is the instantaneous time-dependent system temperature  calculated during the simulation, and $\rm v_{ij}(t)$ is the instantaneous velocity component of the $\rm  i^{th} $ particle. Within the equilibration subroutine, the velocity components of each particle continue to be adjusted using the Equation \ref{eq:scaled} until the time averaged desired system temperature is achieved.\\ \\
A cutoff radius of $\rm r_{c}=2.5$ and the integration time step of  $ \rm dt=0.003$ are used for this test-simulations. We positon the atoms on the lattice sites, whose lengths are equal to $ \rm (1/\rho)^{1/3}$, of simple cubic lattice before starting the equilibration phase. After equilibration, we perform the production run for 20 million simulation time steps. We use the Velocity Verlet integration algorithm \cite{Verlet1967} for all the test-runs as well.\\ \\
\begin{figure}[H]
	\centering
	\includegraphics[height=10cm]{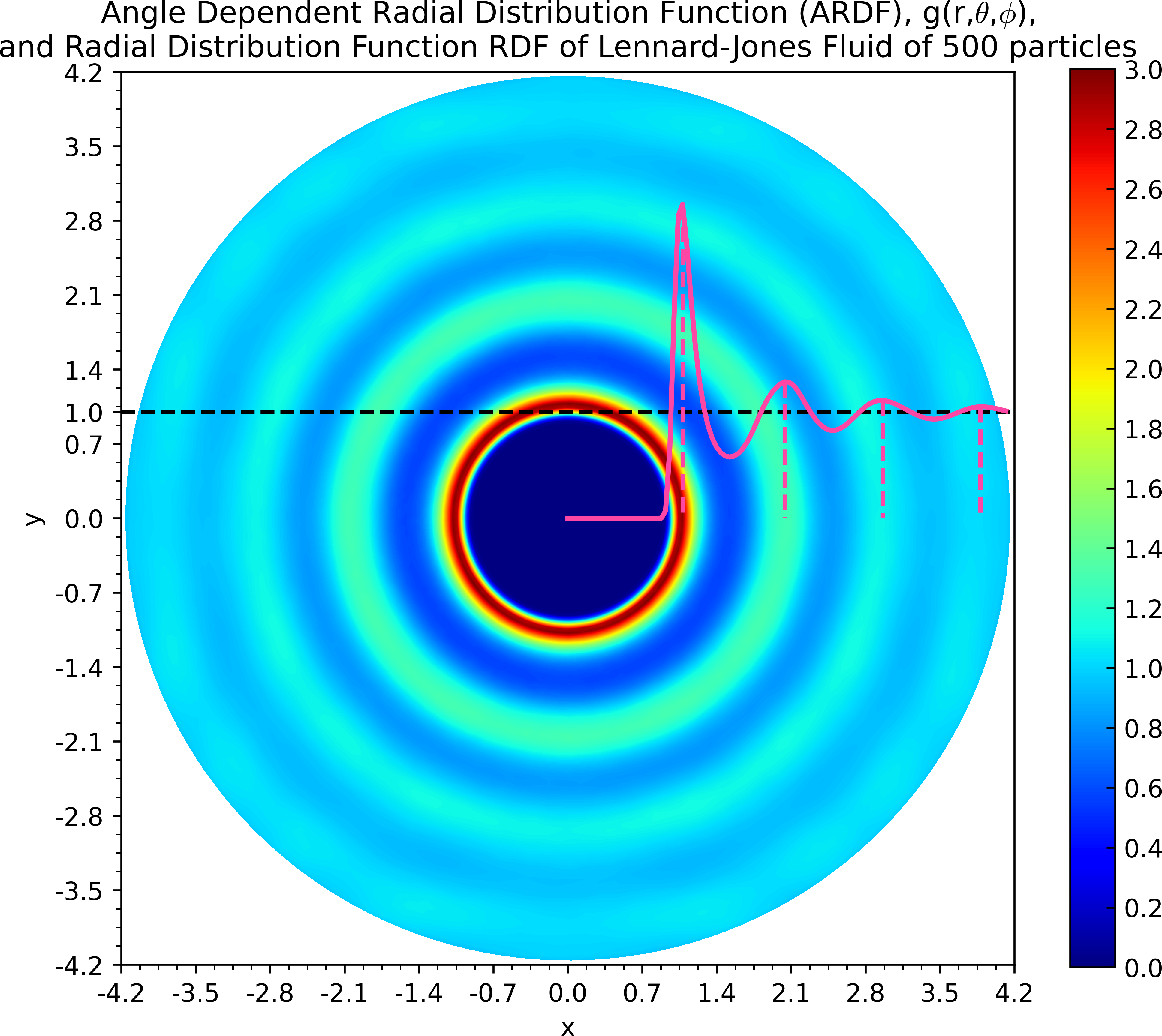}
	\caption{The two-dimensional contour map of the three-dimensional angle dependent radial distribution function ARDF, $ \rm g(r, \theta, \phi)$, of a Lennard-Jones liquid with 500 particles. The first four spherical coordination spheres (concentric annuluses) of decreasing amplitudes for a central liquid-state atom at the origin are seen at the contour map. We also superimposed the one-dimensional radial distribution function RDF (pink) with four crests decreasing amplitudes corresponding to the central intensities of these four annuluses onto the contour map. The RDF is normalized to one (black dashed line). }
	\label{fgr:ardf500LJ}
\end{figure}
\noindent
The one-dimensional radial distribution function was calculated over 20 million
simulation time steps. The mesh resolution in the radial direction is $\rm d_{rad}=0.04$ for the RDF calculation. The two-dimensional contour map of the three-dimensional angle dependent radial distribution function ARDF, $ \rm g(r, \theta, \phi)$, was computed over 4 million simulation time steps. The corresponding mesh resolutions for the radial, $\rm dr $, the azimuthal $\rm d\theta $, and the polar direction $\rm d\phi $ are equal to $ \rm 0.04$ as well. Both the ARDF and the RDF are normalized to one (see Figure \ref{fgr:ardf500LJ}).
\begin{figure}[H]
	\centering
	\includegraphics[height=6cm]{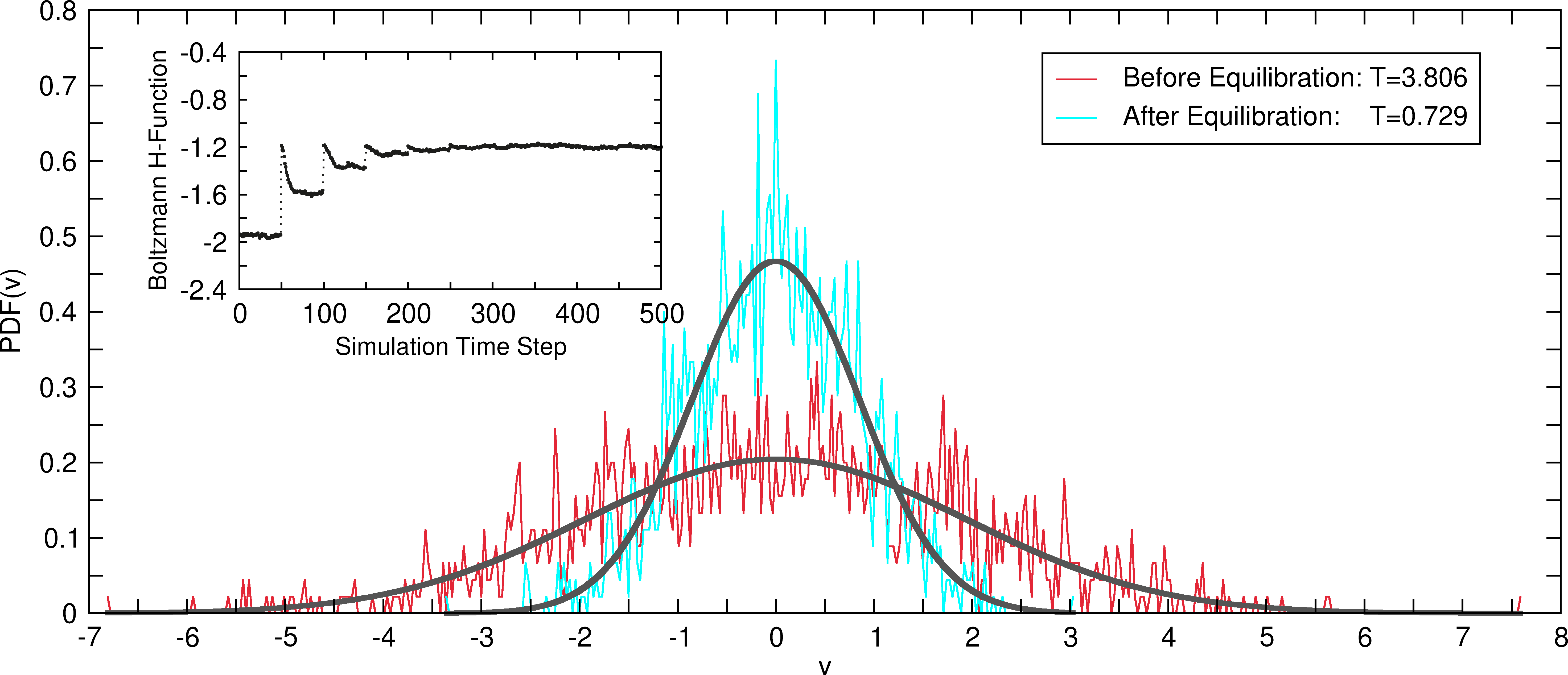}
	\caption{The main figure shows two velocity-dependent probability density functions calculated for only one snapshot as a function of three dimensional velocity components (blue and red colored figures). Black colored solid lines were obtained by using a Gaussian curve fitting. The fitting parameters, $\rm T=0.729 $ and $\rm T=3.806 $, are the dimensionless temperatures after and before the equilibration process, respectively. The inset Figure depicts Boltzmann H-function calculated during equilibration as a function of simulation time step.}
	\label{fgr:MaxwellBoltzmann}
\end{figure}
\noindent
By calculating the velocity distribution function and Boltzmann H-function, we can check whether or not the Lennard-Jones 
liquid of 500 atoms is at the thermodynamic equilibrium. 
We computed Boltzmann H-fuction during the equilibration
process by using the NVE ensemble. 
For the equilibration phase, we utilized the velocity-scaling 
thermostat to reach the desired averaged system temperature, $\rm T=0.722 $. We 
implemented the velocity-scaling procedure at 
every 50 simulation time steps for 20000 simulation time steps in total. 
At the desired thermodynamic equilibrium state, the velocity distribution 
of the particles was observed to obey Maxwellian distribution as shown in the Figure 
\ref{fgr:MaxwellBoltzmann}. At this Figure, we see 
two different probability density function, PDF(v), as a function of velocity components of the particles. Red one belongs to 
the distribution before the equilibration, the blue one belongs to the 
distribution after the equilibration. For this purpose, we used the bin size of 0.03 unit for the calculations of the probability density function and Boltzmann H-function. PDFs were calculated using the three dimensional velocity coordinates at the 20000th time step.
 \begin{equation}
 	\rm PDF(v) = \frac{1}{\sqrt{2\pi T}} e^{-\frac{v^{2}}{2T}}
 	\label{eq:MBD}
 \end{equation}
 \noindent
By using the dimensionless Gaussian distribution function with zero mean velocity in the Equation \ref{eq:MBD}, we did a curve fitting (black colored solid lines) on these two curves as seen in the Figure \ref{fgr:MaxwellBoltzmann}. In Equation \ref{eq:MBD}, variance is equal to the desired dimensionless system temperature T. 
As you can see in the Figure, $\rm T=0.729$ is approximately equal to 0.722 as we wish after the equilibration process. \\  \\
The inset figure in the Figure \ref{fgr:MaxwellBoltzmann} shows Boltzmann H-function versus simulation time step graph. As we said before, we did calculate the Boltzmann H-function during the velocity-scaling. At every 50 simulation time step we scaled the velocities, by doing so, we disturb the system from the one equilibrium state to the another equilibrium state. At the time which the velocities are scaled, we see that Boltzmann H-function increases suddenly and after that it starts to decrease until the system reaches to new equilibrium state as we expect ($\rm dH/dt< 0$, out-of-equilibrium) and then dH/dt=0 again at the equilibrium. As we observe from the inset Figure, the response of the Boltzmann H-function to this abrupt disturbance is a sudden increase of the Boltzmann H-function value. As we see in inset figure in the Figure \ref{fgr:MaxwellBoltzmann}, after approximately 300 simulation time steps the LJ liquid system reaches the thermodynamic equilibrium state at the desired system temperature of T=0.722. 
\section{One Dimensional Radial Distribution Functions for Nanofluids}
\begin{figure}[H]
	\centering
	\includegraphics[height=3cm]{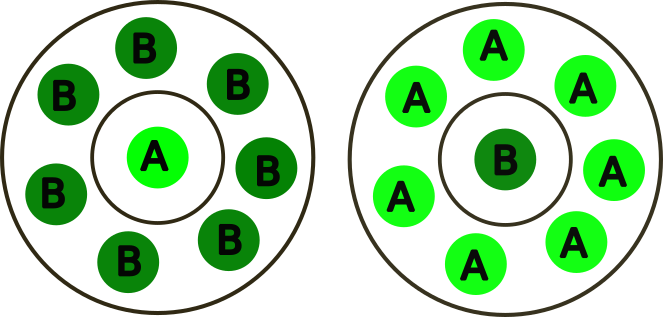}
	\caption{A schematic representation of local particle distributions of B and A atoms around A and B atoms, respectively.}
	\label{fgr:rdfmixture}
\end{figure}

\begin{figure}[H]
	\centering
	\includegraphics[height=4.3cm]{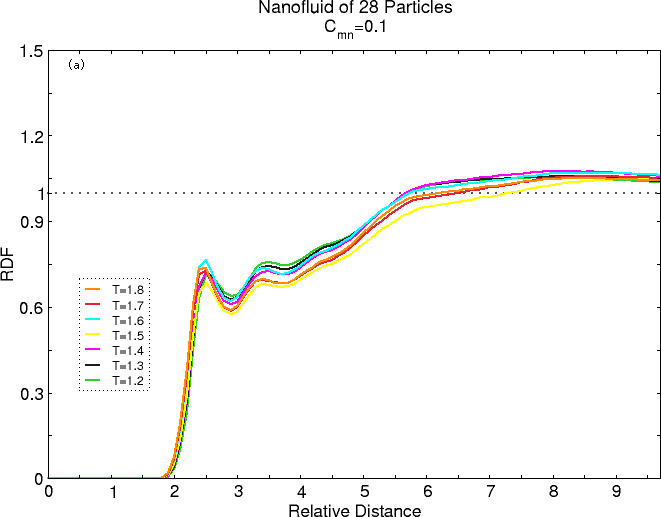}
	\includegraphics[height=4.3cm]{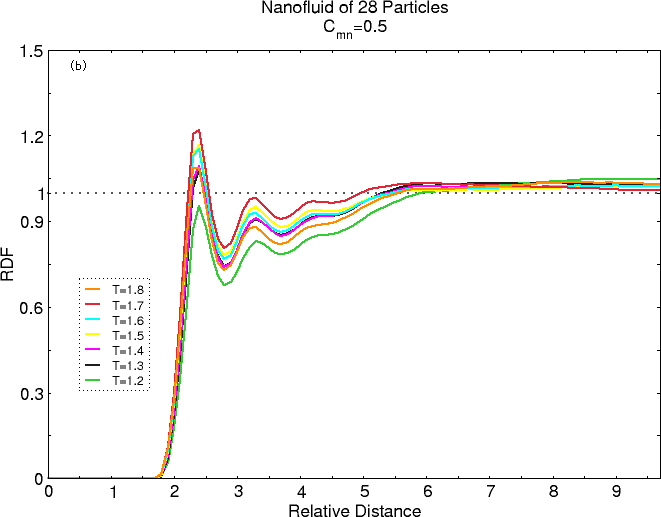}
	\includegraphics[height=4.3cm]{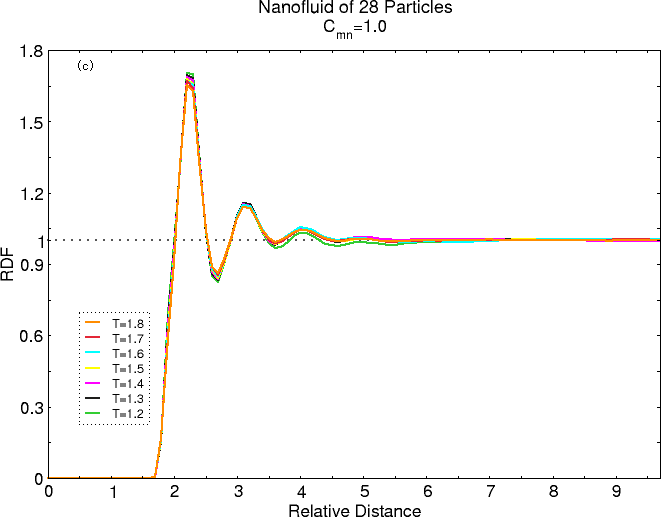}
	\caption{Radial distribution function (RDF) as a function of relative distance for nanofluids of 28 particles. For each $\rm C_{mn}$ value we changed the temperature for seven times.}
	\label{fgr:28pC123T12-18}
\end{figure}

\begin{figure}[H]
	\centering
	\includegraphics[height=4.3cm]{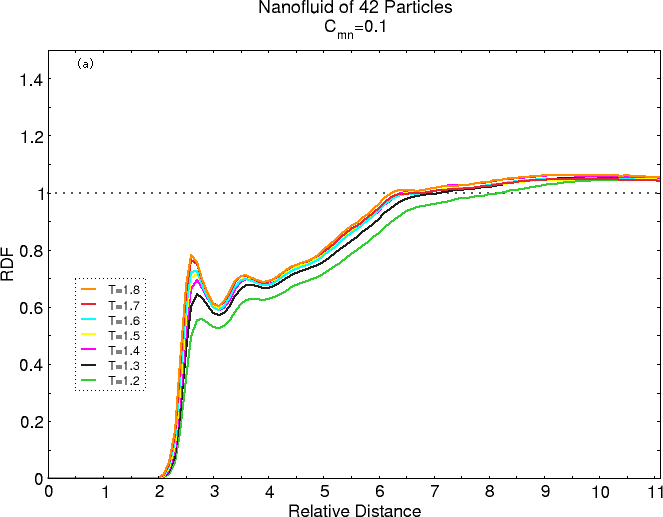}
	\includegraphics[height=4.3cm]{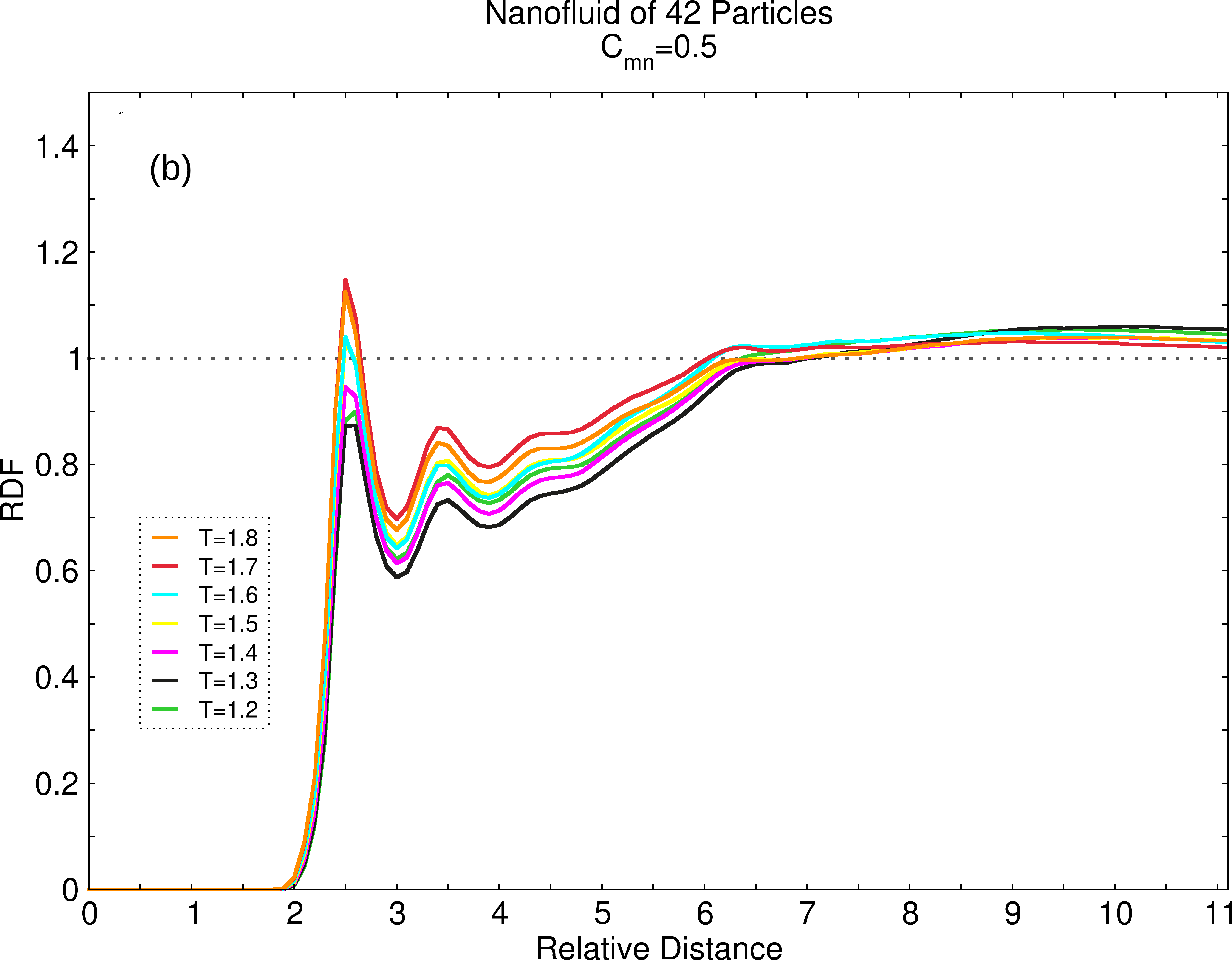}
	\includegraphics[height=4.3cm]{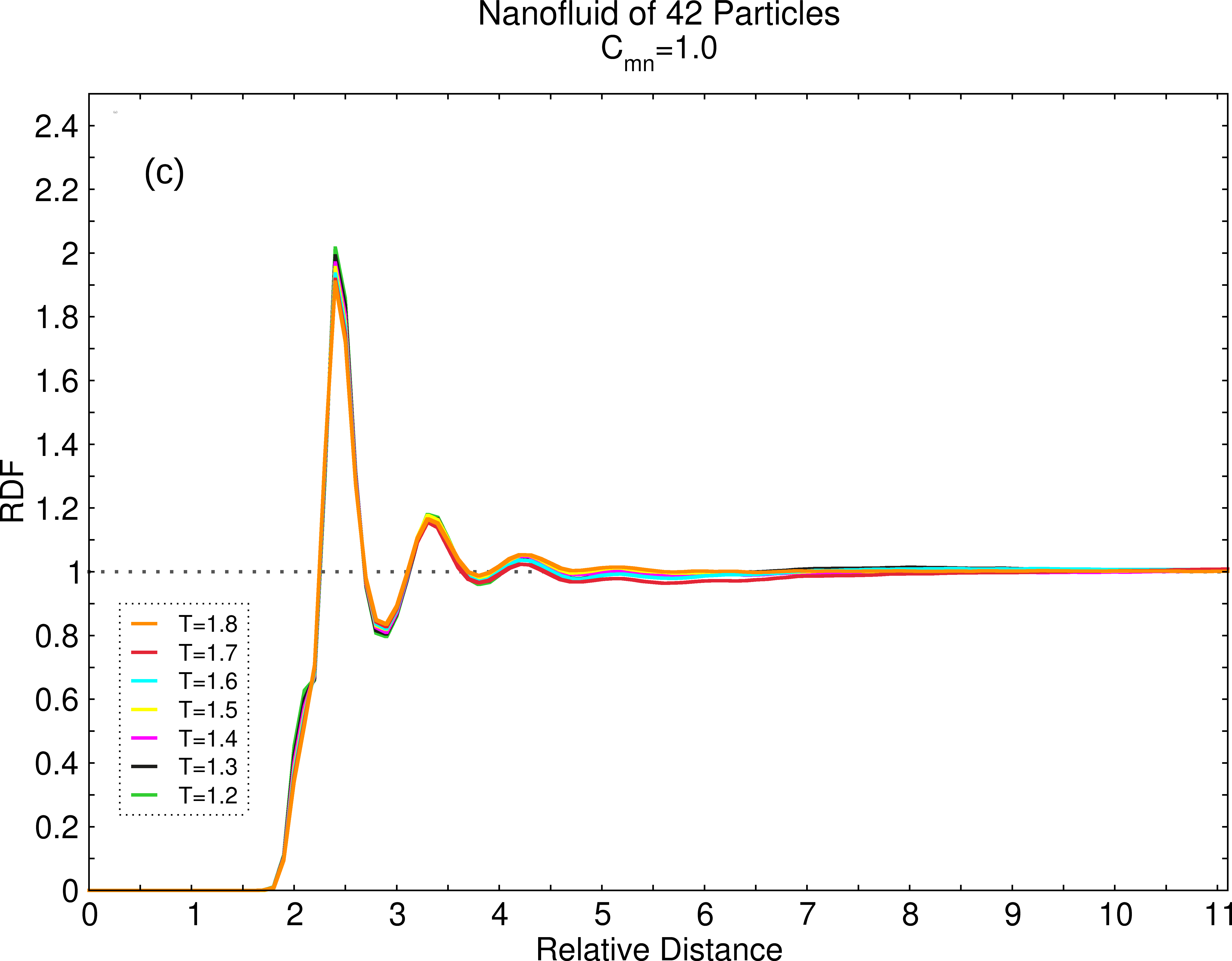}
	\caption{Radial distribution function (RDF) as a function of relative distance for nanofluids of 42 particles. For each $\rm C_{mn}$ value we changed the temperature for seven times.}
	\label{fgr:42pC123T12-18}
\end{figure}

\begin{figure}[H]
	\centering
	\includegraphics[height=4.3cm]{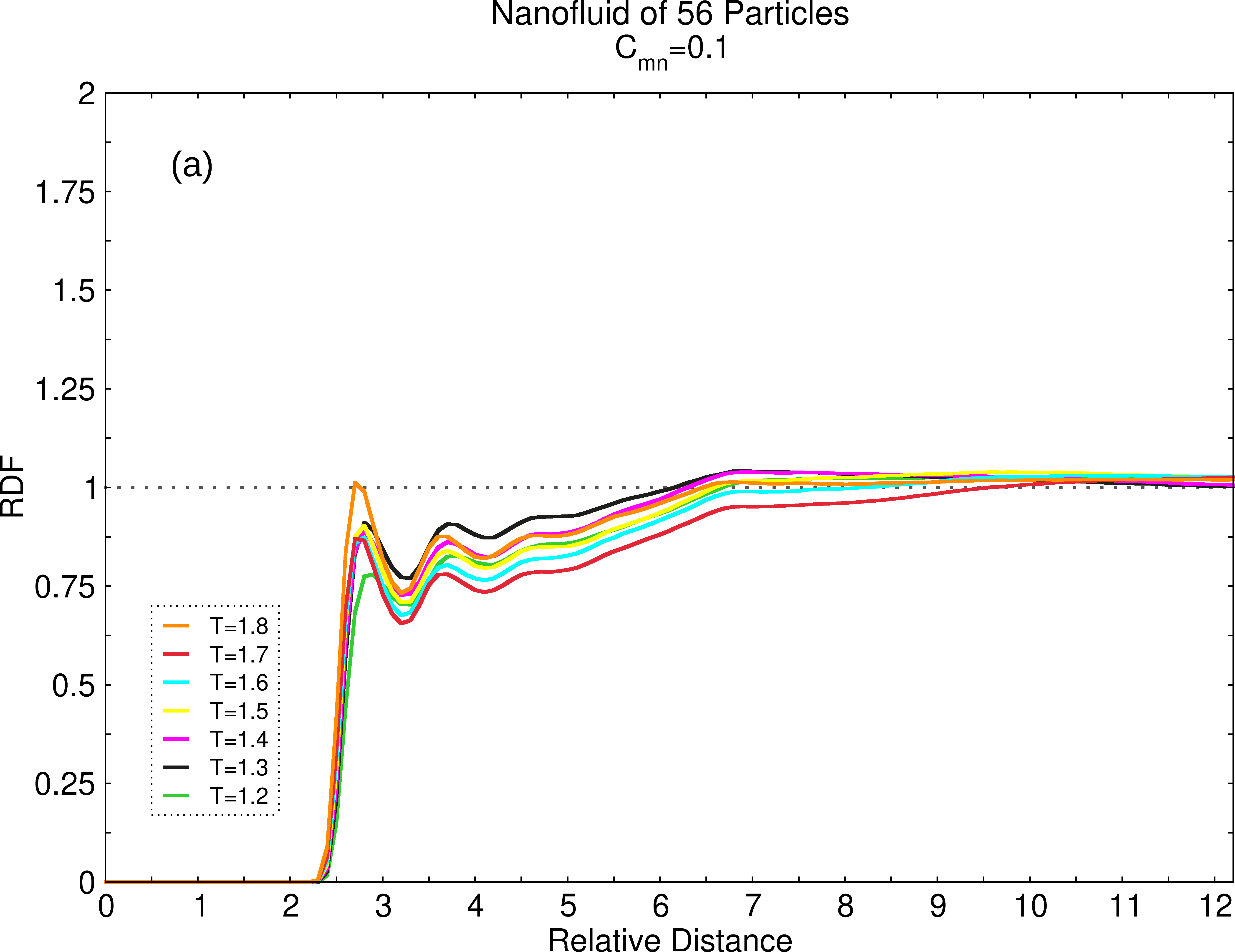}
	\includegraphics[height=4.3cm]{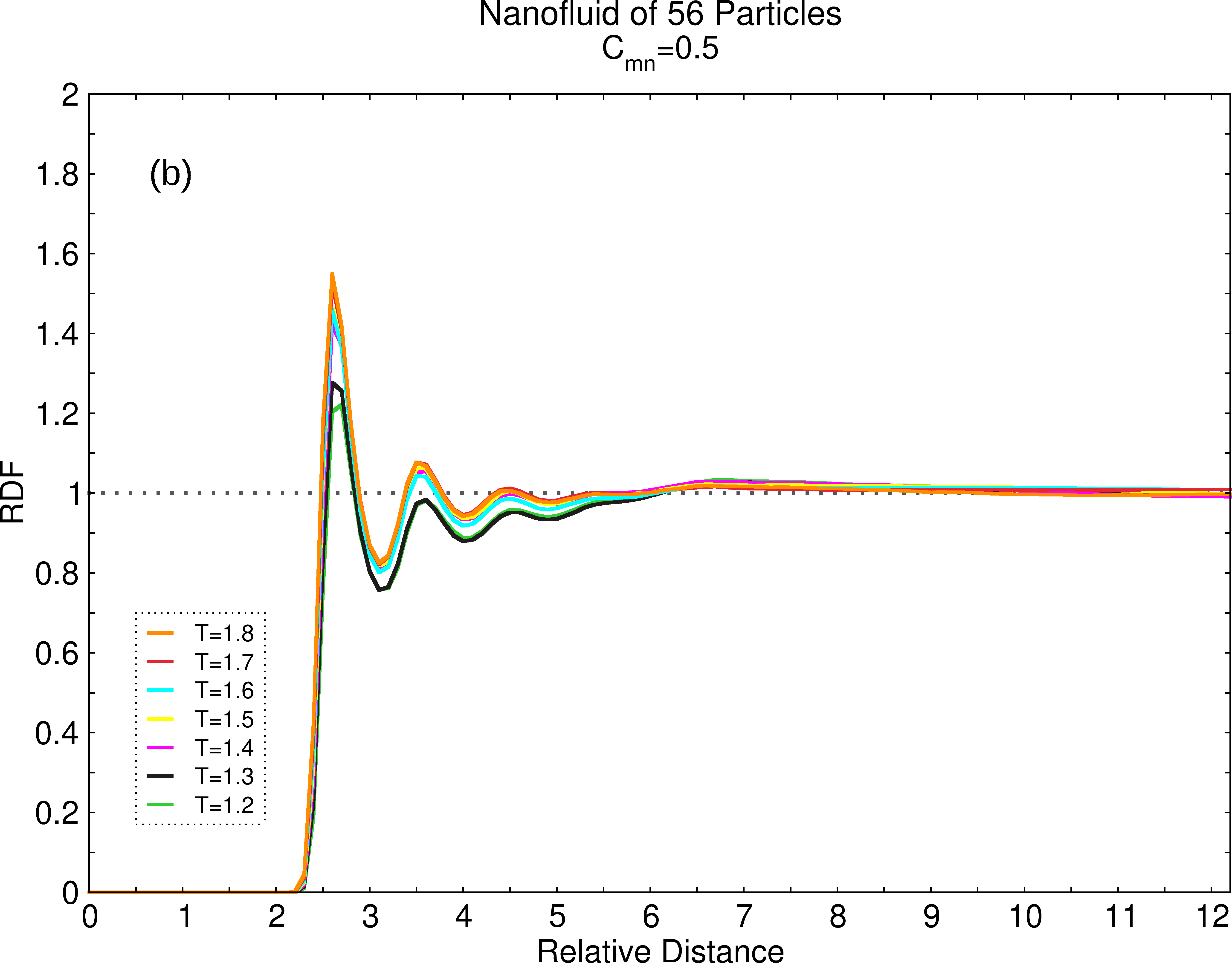}
	\includegraphics[height=4.3cm]{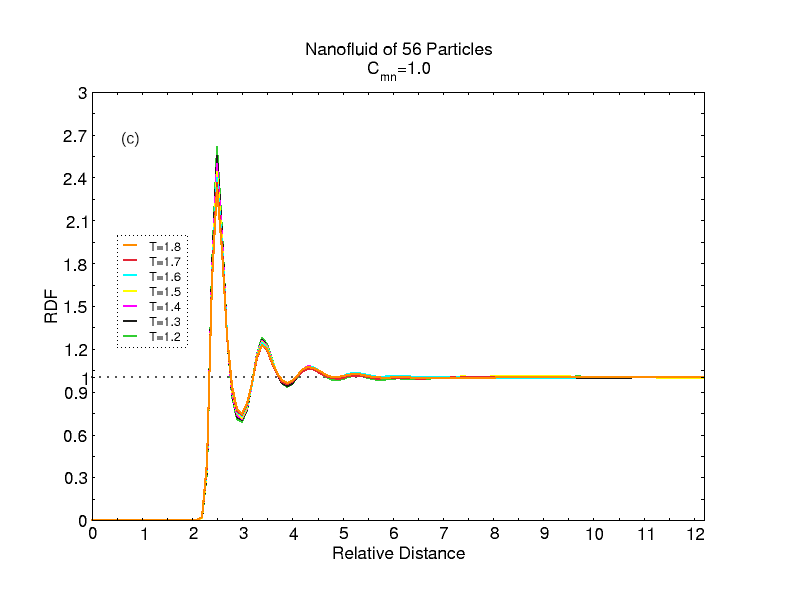}
	\caption{Radial distribution function (RDF) as a function of relative distance for nanofluids of 56 particles. For each $\rm C_{mn}$ value we changed the temperature for seven times.}
		\label{fgr:56pC123T12-18}
\end{figure}

 Systems made up of two or more distinct chemical species are called mixtures. Binary mixes are made up of just two distinct species. In this research, the nanofluids consist of 
nanoparticles of 28, 42, and 56 particles and a polymeric base liquid. The difference in Lennard Jones type-London dispersion bonding capacity between nanoparticle atoms and polymer liquid for different temperature and $\rm C_{mn}$ \cite{Barrat1999} coefficients (See Figure \ref{fgr:interactions} and Eq.\ref{eq:UmCnpa}-\ref{eq:ModfULJ}) can be confirmed by the one-dimensional radial distribution functions. The interaction and physical bonding states of atoms and/or molecules can be examined using the RDFs.\\ \\
In a binary mixture like the nanofluids used in this paper, one can calculate four different types RDFs such as $ \rm g_{AA},\; g_{BB},\; g_{AB}, \; g_{BA}, \; g_{all,all}$, in here $ \rm g_{AB}= g_{BA} $. For the binary mixture number density $\rm \rho$: 
\begin{equation}
	\rm \rho=(N_{A}+N_{B})/V_{sim}  \Longleftrightarrow  \rm \rho=N_{A}/V_{sim}+N_{B}/V_{sim} =\rho_{ A}+\rho_{ B}
\end{equation}
where $ \rm \rho_{ A}$ and $ \rm \rho_{ B}$ number densities of A and B atoms, respectively. $\rm N_{A} $ and $\rm N_{B} $ are also the total number of A and B type particles in the mixture, respectively.
\noindent
if we want to investigate 
the distribution of B particles around A particles, we need to calculate the local radial distribution function $ \rm g_{AB}$:\\ \\
\begin{equation}
	\rm g_{AB}(\Delta r_{n})= \rm \frac{\langle\rho_{B}(\Delta r_{n})\rangle}{\rho_{B}}
\end{equation}
\begin{equation}
	\rm \langle\rho_{B}(\Delta r_{n})\rangle = \rm \frac{N_{B,total}(\Delta r_{n} )}{V_{shell}(\Delta r_{n} )N_{A}N_{frame}},
\end{equation}
\begin{equation}
	\rm g_{AB}(\Delta r_{n})= \rm \rm \frac{V_{sim}N_{B,total}(\Delta r_{n} )}{N_{B}V_{shell}(\Delta r_{n} )N_{A}N_{frame}}, \label{eq:gab}
\end{equation}
\noindent
 We must compute the local pair distribution function $ \rm g_{BA}$ in order to examine the distribution of A particles around B particles (See the Figure \ref{fgr:rdfmixture}):
 \begin{equation}
 	\rm g_{BA}(\Delta r_{n})= \rm \frac{\langle\rho_{A}(\Delta r_{n})\rangle}{\rho_{A}}
 \end{equation}
 \begin{equation}
 	\rm \langle\rho_{A}(\Delta r_{n})\rangle = \rm \frac{N_{A,total}(\Delta r_{n} )}{V_{shell}(\Delta r_{n} )N_{B}N_{frame}},
 \end{equation}
  \begin{equation}
 	\rm g_{BA}(\Delta r_{n})= \rm \rm \frac{V_{sim}N_{A,total}(\Delta r_{n} )}{N_{A}V_{shell}(\Delta r_{n} )N_{B}N_{frame}}, \label{eq:gba}
 \end{equation}
\noindent
From the Equations \ref{eq:gab} and \ref{eq:gba}, one can easily see that $ \rm g_{AB}(\Delta r_{n})=\rm g_{BA}(\Delta r_{n})$. Also one can find that the local pair correlation function $\rm g_{AA}$, which gives the local distribution of A atoms around the single A particle at the center. And in a similar manner, local B distributions around B atoms, $\rm g_{BB}$,  and local distributions of all types of atoms around each one type atom, $\rm g_{all,all}$, can be calculated as follows: \\ \\
\begin{equation}
\rm g_{AA}(\Delta r_{n})= \rm \rm \frac{V_{sim}N_{A,total}(\Delta r_{n} )}{(N_{A}-1)V_{shell}(\Delta r_{n} )N_{A}N_{frame}},
\end{equation}
\begin{equation}
\rm g_{BB}(\Delta r_{n})= \rm \rm \frac{V_{sim}N_{B,total}(\Delta r_{n} )}{(N_{B}-1)V_{shell}(\Delta r_{n} )N_{B}N_{frame}},
\end{equation}
\begin{equation}
\rm g_{all,all}(\Delta r_{n})= \rm \rm \frac{V_{sim}N_{all,total}(\Delta r_{n} )}{(N_{all}-1)V_{shell}(\Delta r_{n} )N_{all}N_{frame}},
\end{equation}
\noindent
We compute the pair correlation functions for nanofluids of three different nanoparticle sizes: 28, 42, and 56 atoms. By changing the coefficient $\rm C_{mn}$ and the system temperature we calculated one dimensional radial distribution functions for three different nanofluids. In Figure \ref{fgr:28pC123T12-18} (a)-(c), we depict RDFs as a function of relative distance for three  $ \rm C_{mn}=0.1, \;0.5,\; 1.0$ values for nanoparticles of 28 atoms. For each $ \rm C_{mn}$, we change the system temperature values to T=1.2, 1.3, 1.4, 1.5, 1.6, 1.7, and 1.8, resulting in seven different simulations. We did the same calculations for nanofluids of 42 and 56 particles as shown in the Figures \ref{fgr:42pC123T12-18} (a)-(c) and the Figures \ref{fgr:56pC123T12-18} (a)-(c), respectively. We performed a total of 72 simulations for three nanoparticles sizes, three distinct  $ \rm C_{mn}$, and seven different temperature values. 
All of the RDFs are normalized to one as seen in the Figures \ref{fgr:28pC123T12-18}-\ref{fgr:56pC123T12-18}. The RDF data were analyzed using a bin size of 0.1. The three dimensional particle coordinates are collected for 500000 frames at every simulation time step. The system pressure is set to zero for $\rm C_{mn}=1.0 $ and the temperature T=1.2 value for nanofluids of 28, 42, and 56 atoms with the number densities at this pressure $ \rm \rho=0.815,\; 0.820,\; and \; 0.825$, respectively.
 \noindent
In the Figure \ref{fgr:28pC123T12-18}, when we decrease the $\rm C_{mn}$ value from 1.0 to 0.1, the affinity between the dispersed phase and the fluid medium decreases drastically, as one expect, since the attractive force term in the Equation \ref{eq:ModfULJ} decreases. The intensity of the first neighborhood peak falls under the one in Figure \ref{fgr:28pC123T12-18} (a). Why is this so? The abscissa of the first peak of the RDF of the nanoparticle with 28 particles for $\rm C_{mn}=0.1$ is greater than the relative distance for $\rm C_{mn}=1.0$, so $\rm V_{shell}(\Delta r_{n} )$ is bigger for $\rm C_{mn}=0.1$ at the corresponding relative distance. Therefore, we understand that the corresponding local total particle number, $\rm N_{total}(\Delta r_{n} )$, at this point must be much less than the one at the relative distance for $\rm C_{mn}=1.0$. This gives rise to a lower peak intensity and a lower local radial distribution, $\rm g(\Delta r_{n} )$, at the same base fluid number density in both cases. The system pressure also increases for $\rm C_{mn}=0.1$ value in comparison to  $\rm C_{mn}=0.5$ and $\rm C_{mn}=1.0$ at the same system volume and temperature conditions. These more weak interactions induced by $\rm C_{mn}$ between the nanoparticles and polymer liquid give rise to an overall pressure increase in the simulation box and accordingly, corresponding increment in the local base fluid number density in practice over against theoretical number density $ \rm \rho=0.741\; (V_{sim}=7558)$, that is to say, the total number of monomers per unit spherical volume around the nanoparticles falls below $\rm \rho=N_{base}/V_{sim}=0.741, \; N_{base}=5600$, which is the total monomer number in the nanofluid with 20 nanoparticles, each containing 28 atoms. However, when we are far away from the central atom of a nanoparticle, due to this increase of the local number density of the base fluid, local radial distribution, $\rm g(\Delta r_{n} )$, does not converge to one, but to a slightly higher value than it as seen in the Figures \ref{fgr:28pC123T12-18} (a)-(b) and the Figures \ref{fgr:42pC123T12-18} (a)-(b). 
\noindent
The lower affinity, or the lower $\rm C_{mn}$ coefficient causes the compression of the polymer sea among the nanoparticles, that is to say, shrinking in the volume occupied by the polymer matrix in the simulation box. In Figures \ref{fgr:56pC123T12-18} (a)-(b), we cannot see the same convergence behavior. For the nanoparticle with 56 atoms, we see that $\rm g(\Delta r_{n} )$ converges to one at a large distance from the central nanoparticle atom. At the system temperature T=1.2 and for $\rm C_{mn}=0.1$ coefficient, the system pressure was measured P=0.7 for 28-atom nanoparticle with the bulk number density $ \rm \rho=0.815$ and P=0.6 for 56-atom nanoparticle with the bulk number density $ \rm \rho=0.825$, so the difference in bulk densities are $\rm 1 \% $.  Therefore, this difference in the bulk densities gives rise to less pressure increase as we decrease the $\rm C_{mn}$ value to 0.1 for 56-atom nanoparticle bulk system, since the system is slighlty more dense $\rm (1 \%) $. This lower pressure increase in this bulk system causes slightly less local average number density increase relative to $ \rm \rho=0.825$ and the volume occupied by the base fluid shrinks a little in the simulation box, and hence when we move away from the central atom of the nanoparticle, RDF directly converges to one as seen in the Figure \ref{fgr:56pC123T12-18} (a)-(b) in comparison to Figure \ref{fgr:28pC123T12-18} (a)-(b) and Figure \ref{fgr:42pC123T12-18} (a)-(b).\\ \\
\noindent
\section{Angle Dependent Three-Dimensional Radial Distribution Functions for Nanofluids}
\begin{figure}[H]
	\centering
	\includegraphics[height=3.1cm]{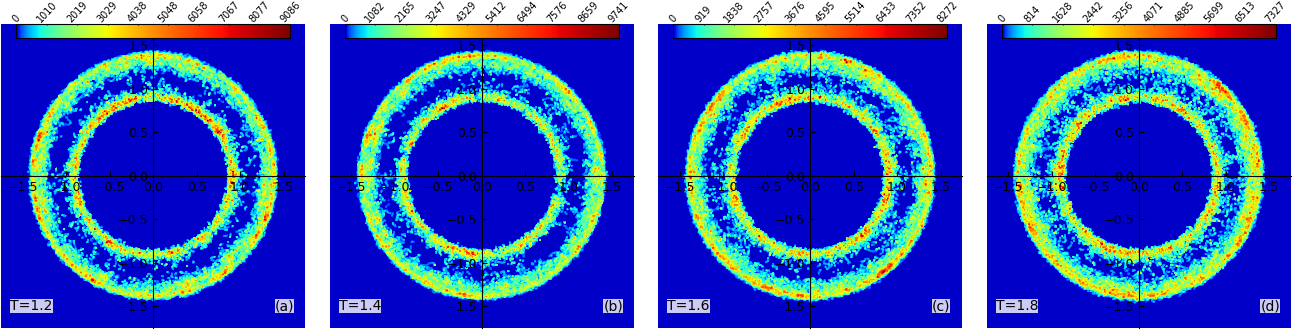}
	\caption{The two-dimensional contour maps of the angle dependent three-dimensional radial distribution functions ARDF,  $ \rm g(r, \theta, \phi)$, for the nanofluid of 20 nanoparticles with 28 atoms for $\rm C_{mn}=0.1$ and for four different temperature values T=1.2, 1.4, 1.6, and 1.8.}
		\label{fgr:ARDF28pC01}
	\end{figure}

\begin{figure}[H]
		\centering
	\includegraphics[height=3.1cm]{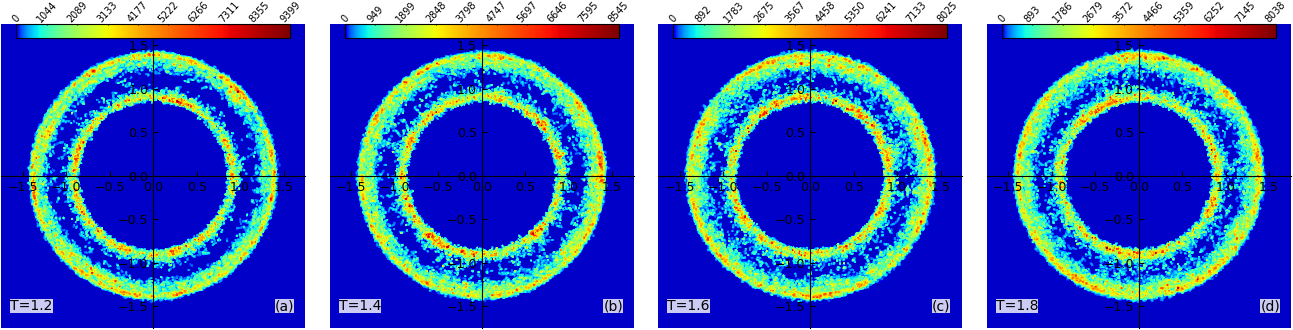}
	\caption{The two-dimensional contour maps of the angle dependent three-dimensional radial distribution functions ARDF,  $ \rm g(r, \theta, \phi)$, for the nanofluid of 20 nanoparticles with 28 atoms for three different $\rm C_{mn}=0.5$ and for four different temperature values T=1.2, 1.4, 1.6, and 1.8.}
		\label{fgr:ARDF28pC05}
	\end{figure}

\begin{figure}[H]
		\centering
	\includegraphics[height=3.1cm]{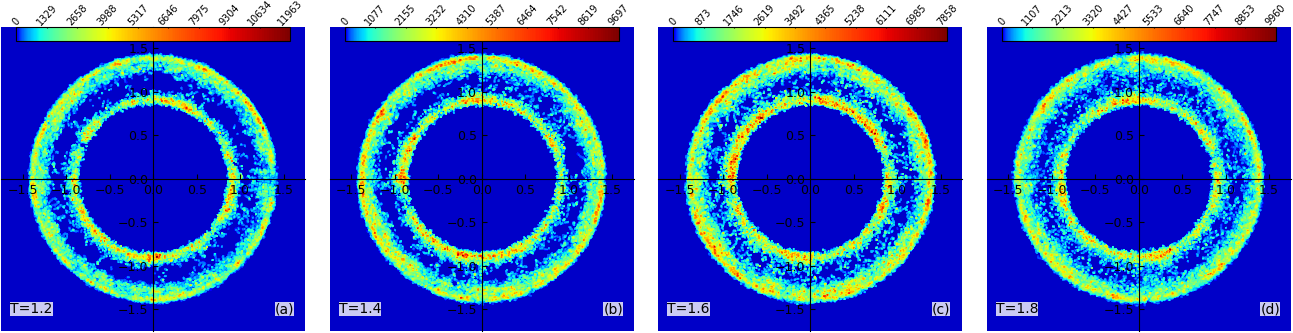}
		\caption{The two-dimensional contour maps of the angle dependent three-dimensional radial distribution functions ARDF,  $ \rm g(r, \theta, \phi)$, for the nanofluid of 20 nanoparticles with 28 atoms for three different $\rm C_{mn}=1.0$ and for four different temperature values T=1.2, 1.4, 1.6, and 1.8.}
	\label{fgr:ARDF28pC10}
\end{figure}

\begin{figure}[H]
	\centering
	\includegraphics[height=3.1cm]{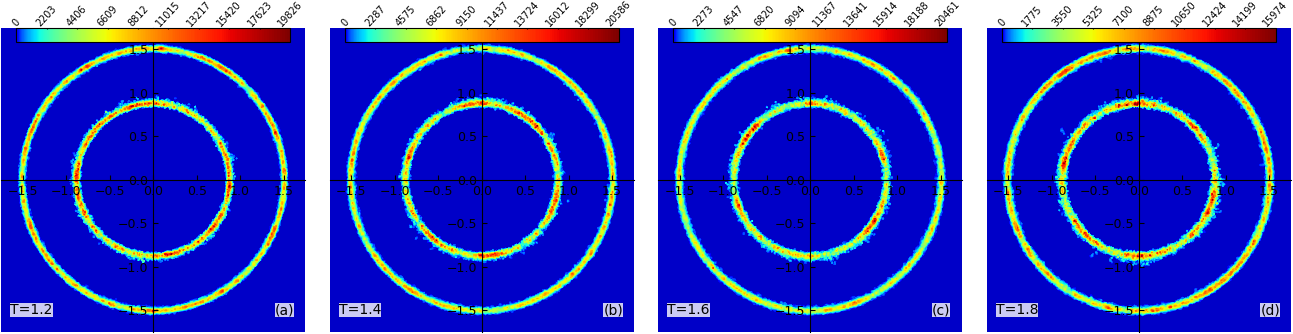}
	\caption{The two-dimensional contour maps of the angle dependent three-dimensional radial distribution functions ARDF,  $ \rm g(r, \theta, \phi)$, for the nanofluid of nanoparticles with 42 atoms for $\rm C_{mn}=0.1$ and for four different temperature values T=1.2, 1.4, 1.6, 1.8.}
	\label{fgr:ARDF42pC01}
\end{figure}

\begin{figure}[H]
	\centering
	\includegraphics[height=3.1cm]{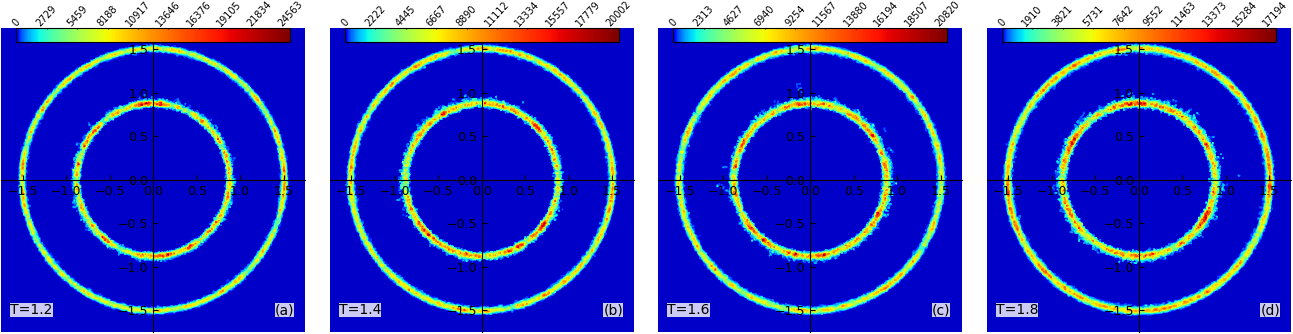}
	\caption{The two-dimensional contour maps of the angle dependent three-dimensional radial distribution functions ARDF,  $ \rm g(r, \theta, \phi)$, for the nanofluid of nanoparticles with 42 atoms for three different $\rm C_{mn}=0.5$ and for four different temperature values T=1.2, 1.4, 1.6, and 1.8.}
	\label{fgr:ARDF42pC05}
\end{figure}

\begin{figure}[H]
	\centering
	\includegraphics[height=3.1cm]{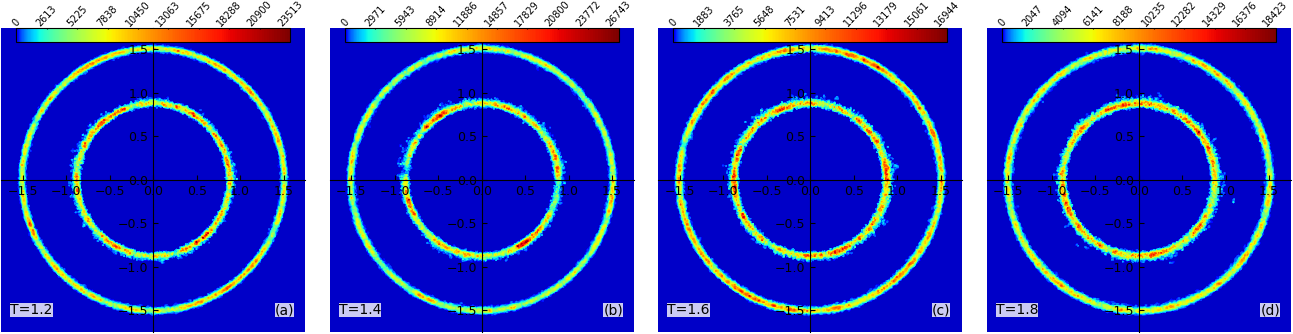}
	\caption{The two-dimensional contour maps of the angle dependent three-dimensional radial distribution functions ARDF,  $ \rm g(r, \theta, \phi)$, for the nanofluid of nanoparticles with 42 atoms for three different $\rm C_{mn}=1.0$ and for four different temperature values T=1.2, 1.4, 1.6, and 1.8.}
	\label{fgr:ARDF42pC10}
\end{figure}

\begin{figure}[H]
	\centering
	\includegraphics[height=3.1cm]{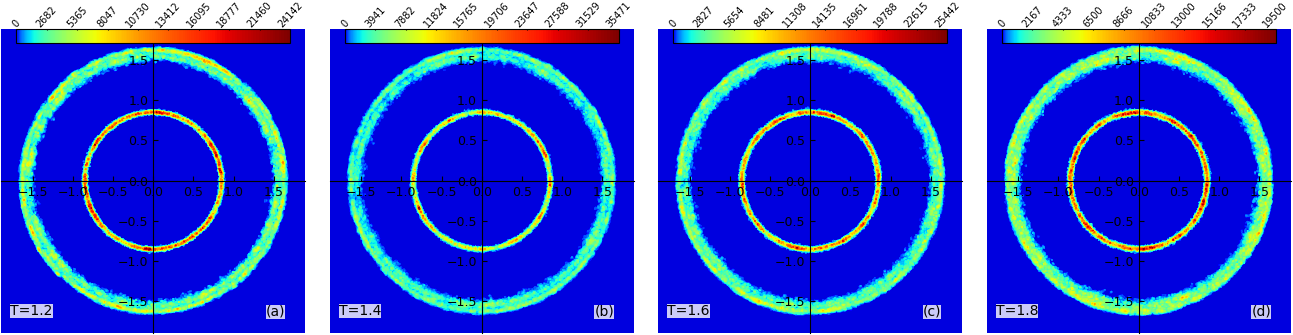}
	\caption{The two-dimensional contour maps of the angle dependent three-dimensional radial distribution functions ARDF,  $ \rm g(r, \theta, \phi)$, for the nanofluid of nanoparticles with 56 atoms for $\rm C_{mn}=0.1$ and for four different temperature values T=1.2, 1.4, 1.6, and 1.8.}
	\label{fgr:ARDF56pC01}
\end{figure}

\begin{figure}[H]
	\centering
	\includegraphics[height=3.1cm]{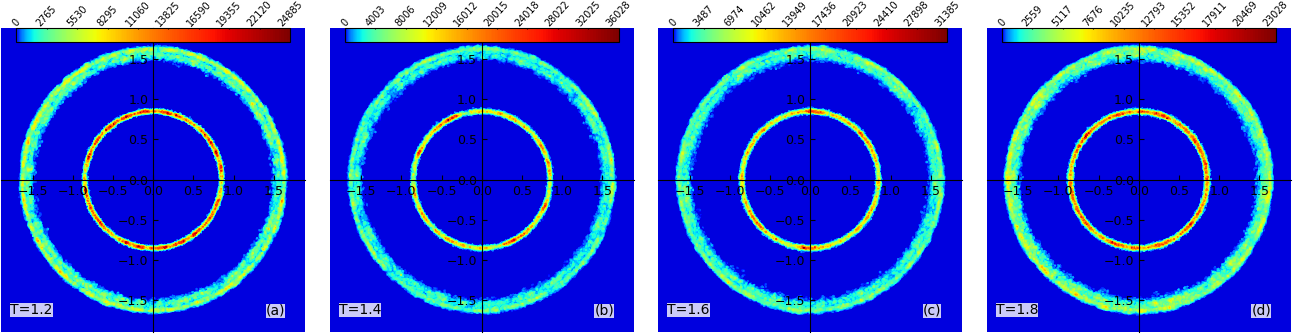}
	\caption{The two-dimensional contour maps of the angle dependent three-dimensional radial distribution functions ARDF,  $ \rm g(r, \theta, \phi)$, for the nanofluid of nanoparticles with 56 atoms for three different $\rm C_{mn}=0.5$ and for four different temperature values T=1.2, 1.4, 1.6, and 1.8.}
	\label{fgr:ARDF56pC05}
\end{figure}

\begin{figure}[H]
	\centering
	\includegraphics[height=3.1cm]{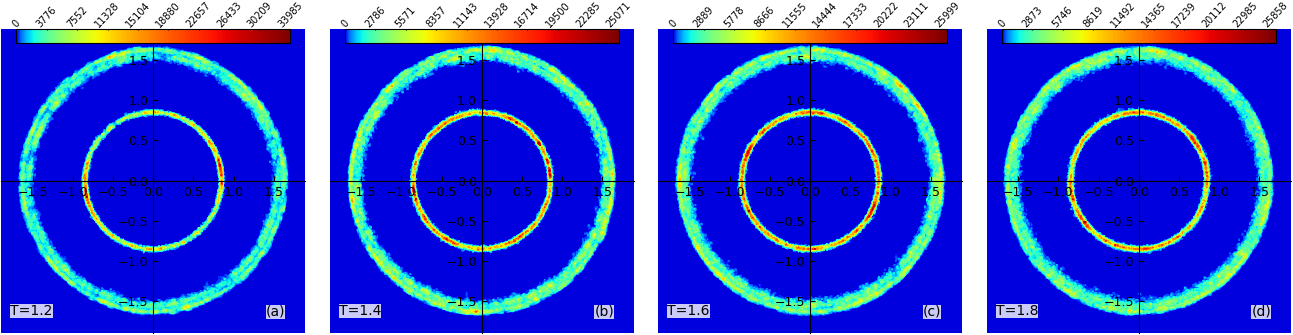}
	\caption{The two-dimensional contour maps of the angle dependent three-dimensional radial distribution functions ARDF,  $ \rm g(r, \theta, \phi)$, for the nanofluid of nanoparticles with 56 atoms for three different $\rm C_{mn}=1.0$ and for four different temperature values T=1.2, 1.4, 1.6, and 1.8.}
	\label{fgr:ARDF56pC10}
\end{figure}

By using the homemade algorithm that we gave in the Section of \textit{"Equilibration of a Nanoparticle"}, we also made the stability analysis for shapes of nanoparticles with three different sizes in a polymer matrix. For this purpose, we calculated the two-dimensional contour maps of the angle dependent three-dimensional radial distribution functions ARDF, $ \rm g(r, \theta, \phi)$ for three different sizes of nanoparticles with 28, 42, and 56 atoms by using this Fortran program. Figures \ref{fgr:ARDF28pC01} (a)-(d), the Figures \ref{fgr:ARDF28pC05} (a)-(d), and the Figures \ref{fgr:ARDF28pC10} (a)-(d) display twelve graphs in total, where each set of four represents 2D contour maps corresponding to $\rm C_{mn}=0.1,\;0.5,\; and \;1.0$ respectively for the nanofluid with 20 nanoparticles, each containing 28 atoms in a polymer matrix containing 5600 monomers as well. Each Figure (a)-(d) within these four-graph sets corresponds to a distinct system temperature of T=1.2, 1.4, 1.6, and 1.8, respectively. For all of the contour maps calculations, we use the mesh resolution of 0.1 unit for the radial, polar, and azimuthal distances. The contour maps were calculated for 500000 simulation time steps in total. As shown by each Figure (a)-(d) within these four-graph sets, the empty space between the inner shell and the outer shell slowly decreases as the system temperature increases. This shows that our nanoparticle of 28 atoms demonstrates the shape instability connected with the system temperature, and it cannot preserve its two discrete concentric spherical shell structure thanks to this temperature effect.
\noindent
Figures \ref{fgr:ARDF42pC01} (a)-(d), the Figures \ref{fgr:ARDF42pC05} (a)-(d), and the Figures \ref{fgr:ARDF42pC10} (a)-(d) display twelve graphs in total, where each set of four represents 2D contour maps corresponding to $\rm C_{mn}=0.1,\;0.5,\; and \;1.0$ respectively for the nanofluid with 20 nanoparticles, each containing 42 atoms in a polymer matrix containing 8400 monomeric units. Each Figure (a)-(d) within these four-graph sets corresponds to a distinct system temperature of T=1.2, 1.4, 1.6, and 1.8, respectively. For all of the 2D contour maps calculations, we use the mesh resolution of 0.1 unit for the radial, polar, and azimuthal distances. The contour maps were calculated for 500000 simulation time steps in total. When we look at these twelve Figures, we see that these nanoparticles of 42 atoms show the shape stability in spite of the temperature increase. Raising the temperature to 1.5 times relative to T=1.2 does not affect the shape of the nanoparticles. Also, the decreasing the affinity between the nanoparticles and the base fluid via the $\rm C_{mn}$ coefficient does not affect the shapes as well. The empty vacuum between two concentric shells, as seen by the Figures, shows the persistency as such in the Figures \ref{fig:RR42} and \ref{fig:42pRDF}.
\noindent
Figures \ref{fgr:ARDF56pC01} (a)-(d), the Figures \ref{fgr:ARDF56pC05} (a)-(d), and the Figures \ref{fgr:ARDF56pC10} (a)-(d) display twelve graphs in total, where each set of four represents 2D contour maps corresponding to $\rm C_{mn}=0.1,\;0.5,\; and \;1.0$ respectively for the nanofluid with 20 nanoparticles, each containing 56 atoms in a polymer matrix containing 11200 monomers. Each Figure (a)-(d) within these four-graph sets corresponds to a distinct system temperature of T=1.2, 1.4, 1.6, and 1.8, respectively. For all of the 2D contour maps calculations, we use the mesh resolution of 0.1 unit for the radial, polar, and azimuthal distances. The contour maps were calculated for 500000 simulation time steps in total. When we look at these twelve Figures, we can see that these nanoparticles of 56 atoms show also the shape stability in spite of the temperature increase too. Raising the temperature to 1.5 times with respect to T=1.2 does not affect the shape of the nanoparticles. Also, the decreasing the affinity between the nanoparticles and the base fluid via the $\rm C_{mn}$ coefficient does not affect the shapes. The empty space between two concentric spherical shells, as seen by the Figures, shows the persistency as such in the Figures \ref{fig:RR56} and \ref{fig:56pRDF}.
\noindent
It appears that the number of atoms making up the nanoparticles
affects the shape stabilitiy and the related two concentric spherical shell nature of them. When we look at the reasons  behind the instability from a holistic perspective, first, one should think of the used FENE radii, $\rm R_{0}=1.52, \;1.62,\; and\; 1.71$ for the nanoparticles of 28, 42, 56 atoms, respectively as an instability factor, and the overall effect of the correspondingly decreased LJ force (intermolecular) intensity on the spherical nanoparticles with increasing FENE radius. The second factor is the rise in the system temperature and the corresponding system pressure increase in the simulation box as well. Thirdly, the utilized cut-off distance, $\rm r_{c}=2\times2^{1/6}$ in the simulations certainly has an impact on discovering in silico stable nanoparticles in terms of shape.
 \section{Mean Force Magnitude Calculations of Nanofluids}
\begin{figure}[H]
	\centering
	\includegraphics[height=3.9cm]{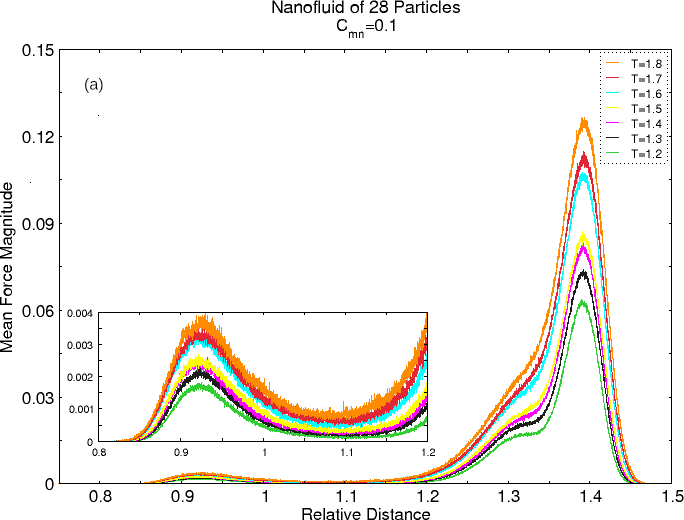}
	\includegraphics[height=3.9cm]{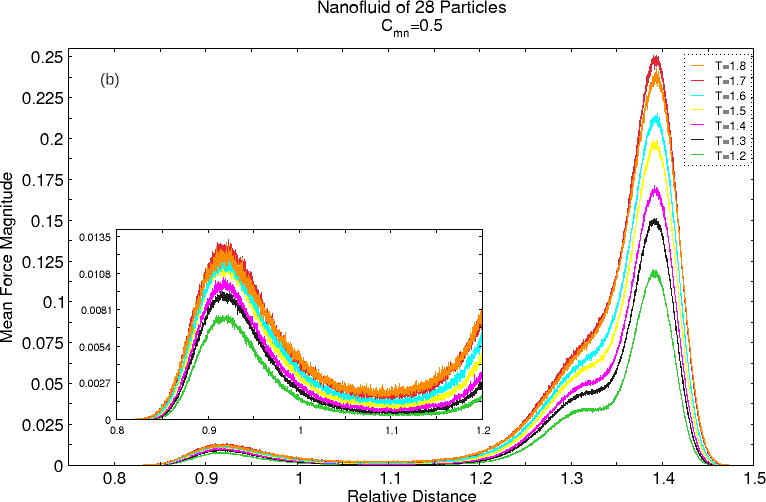}
	\includegraphics[height=3.9cm]{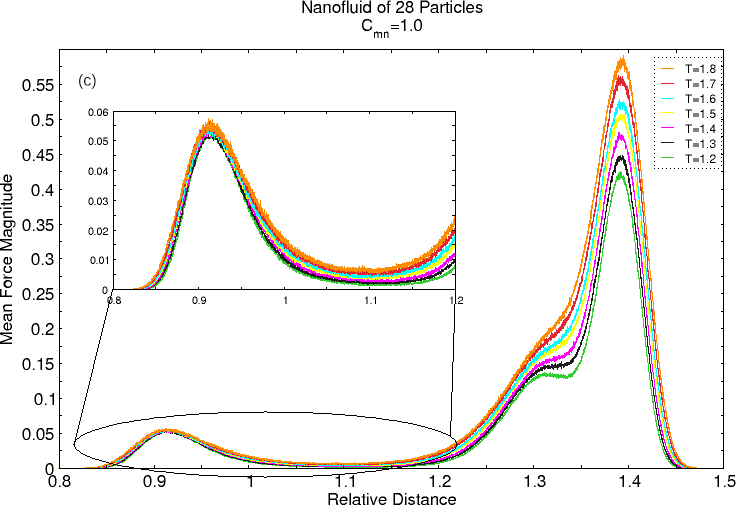}
	\caption{Mean force calculations for the nanofluids with 28 particles for $\rm C_{mn}=0.1,\;0.5,\;and\;1.0$ coefficients in (a), (b), and (c), respectively. In each three Figures we calculated mean force magnitudes for seven distinct system temperature values: T=1.2, 1.3, 1.4, 1.5, 1.6, 1.7, and 1.8.  }
	\label{fgr:meanforcemag28p}
\end{figure}

\begin{figure}[H]
	\centering
	\includegraphics[height=4.2cm]{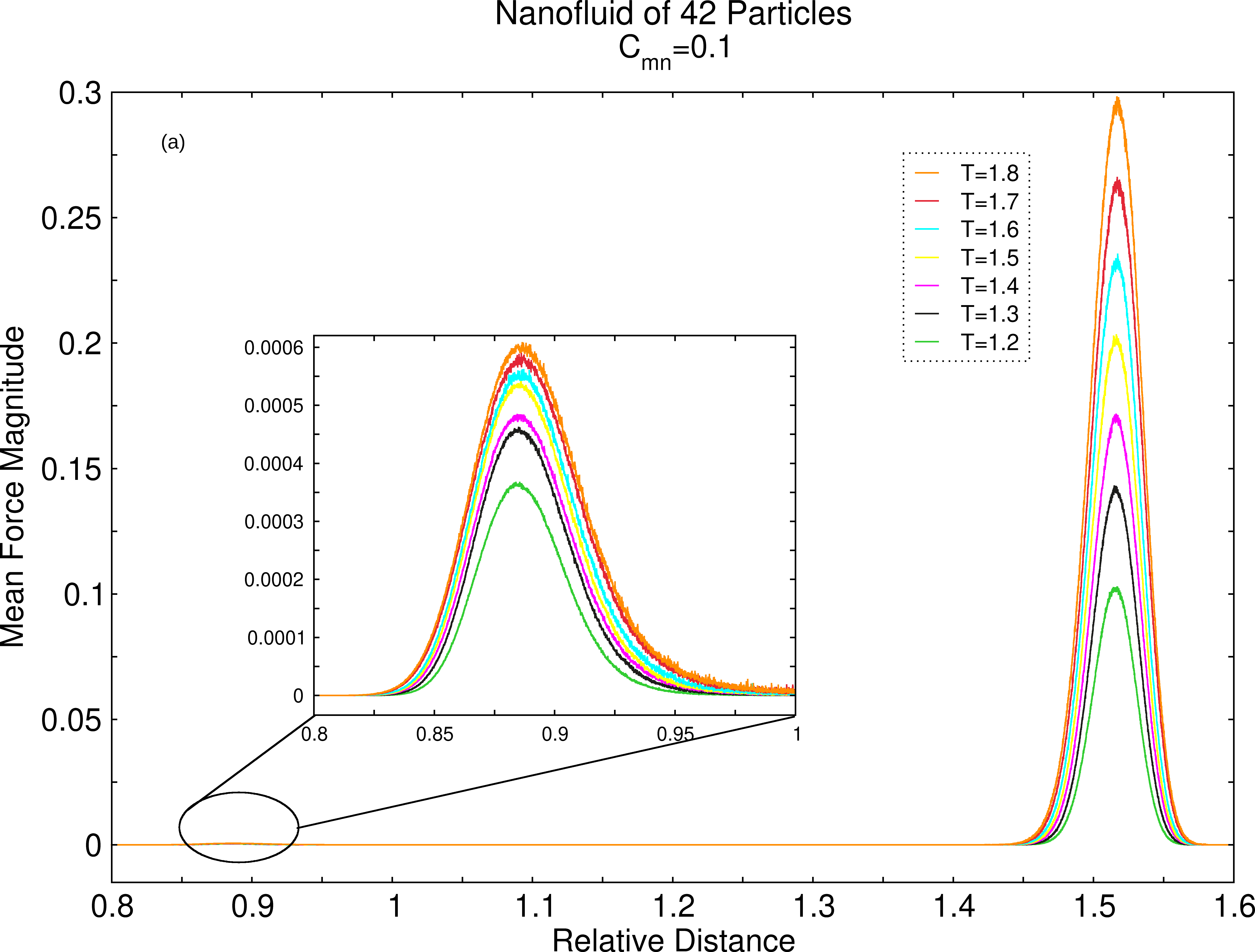}
	\includegraphics[height=4.2cm]{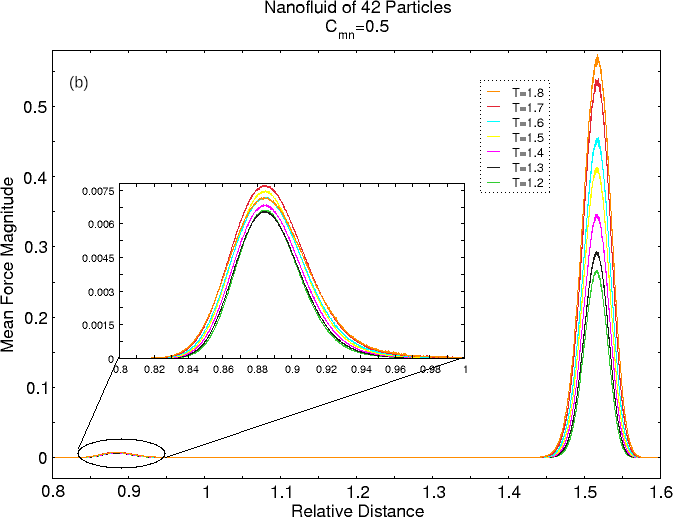}
	\includegraphics[height=4.2cm]{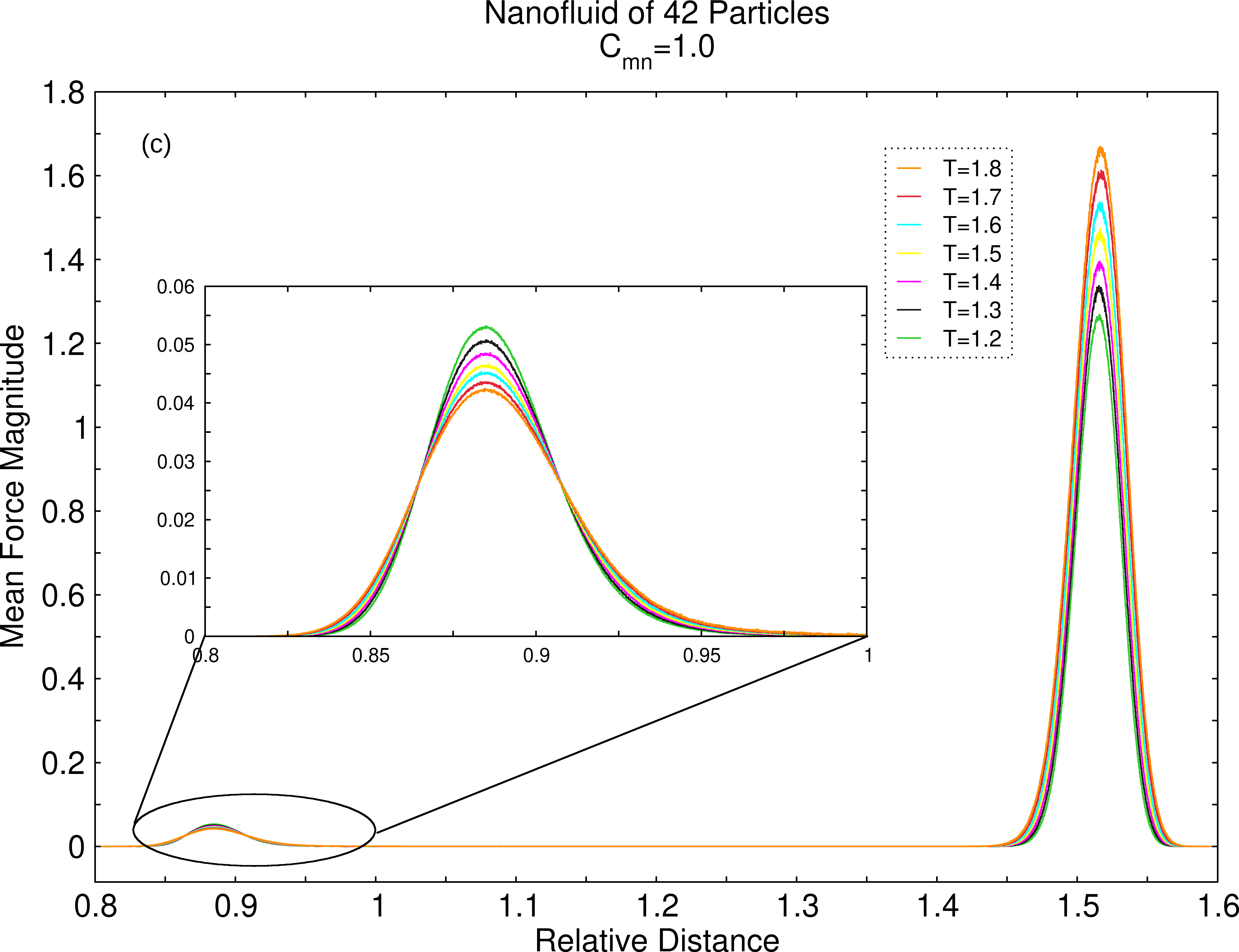}
	\caption{Mean force calculations for the nanofluids with 42 particles for $\rm C_{mn}=0.1,\;0.5,\;and\;1.0$ coefficients in (a), (b), and (c), respectively. In each three Figures we calculated mean force magnitudes for seven distinct system temperature values: T=1.2, 1.3, 1.4, 1.5, 1.6, 1.7, and 1.8.  }
	\label{fgr:meanforcemag42p}
\end{figure}

\begin{figure}[H]
	\centering
	\includegraphics[height=4.2cm]{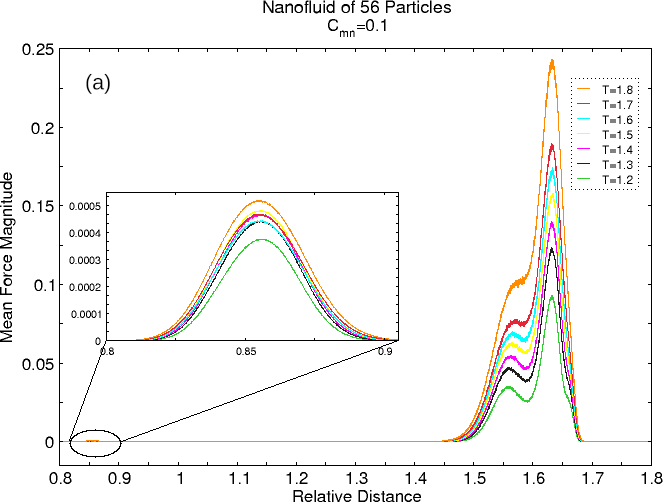}
	\includegraphics[height=4.2cm]{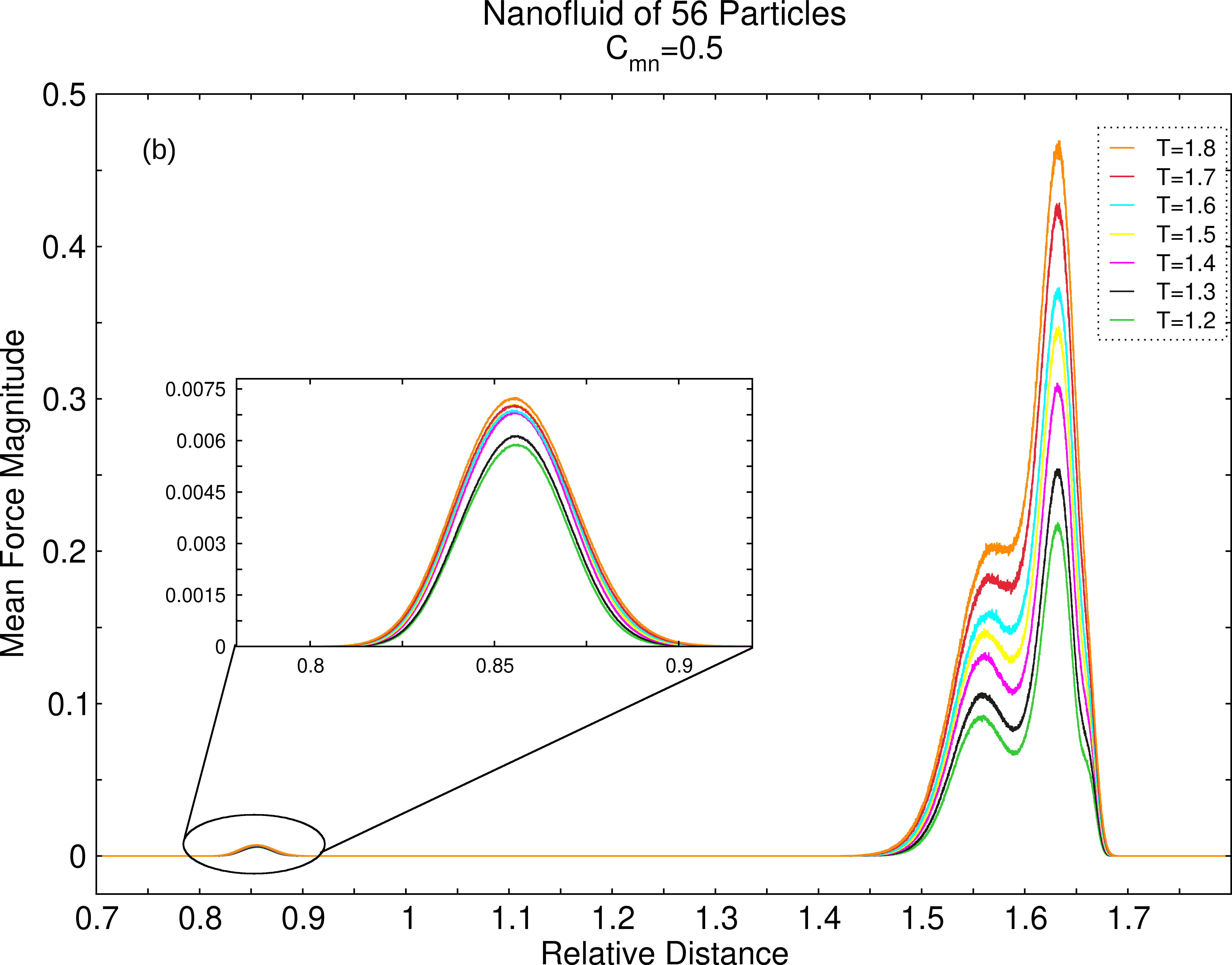}
	\includegraphics[height=4.2cm]{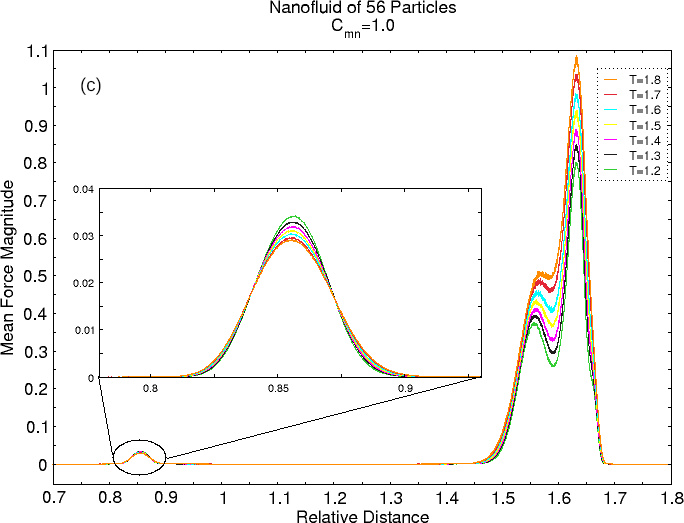}
	\caption{Mean force calculations for the nanofluids with 56 particles for $\rm C_{mn}=0.1,\;0.5,\;and\;1.0$ coefficients in (a), (b), and (c), respectively. In each three Figures we calculated mean force magnitudes for seven distinct system temperature values: T=1.2, 1.3, 1.4, 1.5, 1.6, 1.7, and 1.8. }
	\label{fgr:meanforcemag56p}
\end{figure}

We also calculated mean force magnitudes exerted on the nanofluids with three different sizes containing 28, 42, and 56 particles for $\rm C_{mn}=0.1,\;0.5,\;and\;1.0$ coefficients and for each $\rm C_{mn}$ by changing the system temperature for seven times, T=1.2, 1.3, 1.4, 1.5, 1.6, 1.7, and 1.8. To the mean force magnitude calculations, we performed the post-processing simulations for 500000 simulation time steps in total by using a bin size of 0.0001 unit for the relative distance. The Figures \ref{fgr:meanforcemag28p} (a)-(c), the Figures \ref{fgr:meanforcemag42p} (a)-(c), and the Figures
\ref{fgr:meanforcemag56p} (a)-(c) depict the mean force magnitude as a function of the relative distance for the nanofluids with the nanoparticles of 28, 42, and 56 atoms, respectively. Each (a), (b), and (c) graphs shows this variation for three distinct $\rm C_{mn}=0.1,\;0.5,\;and\;1.0$ coefficient, respectively. For each $\rm C_{mn}$ value, we changed the system temperature for seven times as well. The average force magnitudes were calculated over 500000 frames and 20 nanoparticle numbers in total. The collected and analysed force vectors represent the Lennard-Jones type intermolecular forces exerted on the nanoparticles inside the polymeric liquid medium. \\ \\
\noindent
The Figure \ref{fgr:meanforcemag28p} (a), (b), and (c)
show, for three different $\rm C_{mn}$ coefficients, the mean force magnitude as a function of relative distance, which is calculated with respect to the central atom of each 20 nanoparticles in the liquid medium. As seen by these three figures, when we increase the affinity between the nanoparticles and the polymeric liquid medium, the time frame and the particle number averaged mean of the magnitude of the net force inreases as we expected. And for each $\rm C_{mn}$ value, if we increase the system temperature, the mean force magnitude also increases too. In these three figures (a)-(c) in Figure \ref{fgr:meanforcemag28p}, the distributions exhibit two different peaks, consisting of a smaller minor peak and a larger major peak. At the same time, the large peaks at the right hand side also consist of two different crests, a smaller one and a bigger one as well. The bimodal distributions at the right hand side peaks result from the allocations of the nanoparticle's atoms in the outer spherical shell. We observe that 
the outer spherical shell of the nanoparticle with 28 atoms consists of at least two adjacent concentric rings, that is to say, the outer shell is more thicker than the inner shell as seen by the Figures \ref{fgr:ARDF28pC01}-\ref{fgr:ARDF28pC10}. This discrepancy between the inner and the outer spherical shells gives rise to the bimodal distribution at the right-hand side dispersions. Looking at the Figures \ref{fgr:meanforcemag28p}(a)-(c), the Figures \ref{fgr:meanforcemag42p} (a)-(c), and the Figures \ref{fgr:meanforcemag56p} (a)-(c) as a whole, we see that they also have the bimodal distributions thanks to the presence of the inner and the outer spherical shells. For instance, in the insets of the Figure \ref{fgr:meanforcemag28p} (a)-(c), we can see the existence of the inner shells as seen by the other Figures \ref{fgr:meanforcemag42p} (a)-(c) and Figures \ref{fgr:meanforcemag56p} (a)-(c). 
\noindent
Looking at the Figures \ref{fgr:meanforcemag42p} (a)-(c) as a whole, we observe that the graphs exhibit the 
bimodal dispersions too. However, we can see that the distributions at the right hand side display 
unimodal mean force distirbutions as a function of the relative distance. This difference results from the dispersion patterns of the atoms of the nanoparticle in the outer spherical shell. As seen by the Figures \ref{fgr:ARDF42pC01} (a)-(d), Figures \ref{fgr:ARDF42pC05} (a)-(d), and Figures \ref{fgr:ARDF42pC10} (a)-(d), the outer shells of the nanoparticle with 42 atoms are thinner than the outer shells of the other nanoparticles with 28 and 56 atoms. 
\noindent
Looking at the Figures \ref{fgr:meanforcemag56p} (a)-(c) as a whole, we observe that the graphs exhibit the 
bimodal dispersions as well. And also, we can see that the distributions at the right hand side display 
bimodal mean force distirbutions as a function of the relative distance too. This difference results from the dispersion patterns of the atoms of the nanoparticle in the outer spherical shell. As seen by the Figures \ref{fgr:ARDF56pC01} (a)-(d), the Figures \ref{fgr:ARDF56pC05} (a)-(d), and the Figures \ref{fgr:ARDF56pC10} (a)-(d), the outer shells of the nanoparticle with 56 atoms are thicker than the outer shell of the nanoparticle with 42 atoms. 
\acknowledgments
\textbf{Acknowledgments} \\ \\
This research is supported by the Scientific and Technological Research Council of Türkiye (TÜBİTAK) under the TÜBİTAK 1001 Project No. 122F155.

\end{document}